\documentclass[useAMS,usenatbib]{mn2e}

\usepackage{graphicx}	
\usepackage{amsmath}	
\usepackage{amssymb}	
\usepackage{upgreek}
\newcommand{\MS}{\ifmmode{\,}\else\thinspace\fi{\rm M}\ifmmode_{\odot}\else$_{\odot}$\fi}
\newcommand{\LS}{\ifmmode{\,}\else\thinspace\fi{\rm L}\ifmmode_{\odot}\else$_{\odot}$\fi}
\newcommand{\RS}{\ifmmode{\,}\else\thinspace\fi{\rm R}\ifmmode_{\odot}\else$_{\odot}$\fi}
\newcommand{\Ke}{\ifmmode{\,}\else\thinspace\fi{\rm K}}

\newcommand{\teff}{\ifmmode T_{\rm eff}\else$T_{\rm eff}$\fi}
\newcommand{\fo}{\ifmmode \nu_{\rm 1O}\else$\nu_{\rm 1O}$\fi}
\newcommand{\fx}{\ifmmode \nu_{\rm x}\else$\nu_{\rm x}$\fi}
\newcommand{\fsh}{\ifmmode \nu_{\rm sh}\else$\nu_{\rm sh}$\fi}
\newcommand{\ash}{\ifmmode A_{\rm sh}\else$A_{\rm sh}$\fi}
\newcommand{\ax}{\ifmmode A_{\rm x}\else$A_{\rm x}$\fi}
\newcommand{\Po}{\ifmmode P_{\rm 1O}\else$P_{\rm 1O}$\fi}
\newcommand{\Px}{\ifmmode P_{\rm x}\else$P_{\rm x}$\fi}
\newcommand{\pxpo}{\ifmmode P_{\rm x}/P_{\rm 1O}\else$P_{\rm x}/P_{\rm 1O}$\fi}

\title[Non-radial pulsation in SMC Cepheids]{Non-radial pulsation in first overtone Cepheids of the Small Magellanic Cloud}
\author[Smolec \& \'Sniegowska]{
R. Smolec$^{1}$\thanks{E-mail: smolec@camk.edu.pl}, M. \'Sniegowska$^{2}$
\\
$^{1}$ Nicolaus Copernicus Astronomical Center, Polish Academy of Sciences, ul. Bartycka 18, 00-716 Warszawa, Poland\\
$^{2}$ Warsaw University Observatory, Al. Ujazdowskie 4, 00-478 Warszawa, Poland\\
}

\begin{document}

\date{Accepted . Received ; in original form }

\pagerange{\pageref{firstpage}--\pageref{lastpage}} \pubyear{2015}

\maketitle

\label{firstpage}

\begin{abstract}
We analyse photometry for 138 first overtone Cepheids from the Small Magellanic Cloud, in which Optical Gravitational Lensing Experiment (OGLE) team discovered additional variability with period shorter than first overtone period, and period ratios in the $P/\Po\!\in\!(0.60,\, 0.65)$ range. In the Petersen diagram these stars form three well separated sequences. The additional variability cannot correspond to other radial mode. This form of pulsation is still puzzling. 

We find that amplitude of the additional variability is small, typically 2--4\thinspace per cent of the first overtone amplitude, which corresponds to 2--5 mmag. In some stars we find simultaneously two close periodicities corresponding to two sequences in the Petersen diagram. The most important finding is the detection of power excess at half the frequency of the additional variability (at subharmonic) in 35\thinspace per cent of the analysed stars. Interestingly, power excess at subharmonic frequency is detected mostly for stars of the middle sequence in the Petersen diagram (74\thinspace per cent), incidence rate is much lower for stars of the top sequence (31\thinspace per cent), and phenomenon is not detected for stars of the bottom sequence. The amplitude and/or phase of the additional periodicities strongly vary in time.

Similar form of pulsation is observed in first overtone RR~Lyrae stars. Our results indicate that the nature and cause of this form of pulsation is the same in the two groups of classical pulsators; consequently, a common model explaining this form of pulsation should be searched for. Our results favour the theory of the excitation of non-radial modes of angular degrees 7, 8 and 9, proposed recently by Dziembowski.
\end{abstract}


\section{Introduction}\label{sec:intro}

The majority of classical Cepheids are single-periodic, radial pulsators. More complex pulsation is not rare, however. Double-mode Cepheids pulsating simultaneously in the radial fundamental and in the radial first overtone modes (F+1O) and in the two lowest-order radial overtones (1O+2O) are known for years. These form of pulsation is recognized based on characteristic period ratios of the excited pulsation modes, $P_{\rm 2O}/P_{\rm 1O}\approx 0.80-0.81$ for 1O+2O pulsators and $P_{\rm 1O}/P_{\rm F}\approx 0.715-0.74$ for F+1O pulsators \citep[e.g.][]{ogle_cep_lmc,ogle_cep_smc}. A well known fact is that period ratio depends on metallicity; characteristic values may slightly differ for stars from different stellar systems. The Optical Gravitational Lensing Experiment \citep[OGLE,][]{ogleIII,ogleIV} observations led to the discovery of other forms of multi-periodic pulsation among Cepheids. Triple-mode radial pulsation, F+1O+2O and 1O+2O+3O, was identified \citep{pamtri,ogle_freaks,ogle_cep_smc,ogle_cep_blg,ogleIV_cep_multi}.  A very interesting triple-mode Cepheid, with 1O, 2O and additional longer-period mode, was discovered with {\it CoRoT} \citep{pbw}. Rare and peculiar double-mode pulsations were also discovered, including 1O+3O pulsation with 2O apparently not excited \citep{ogle_freaks,ogleIV_cep_multi} and first double-mode 2O+3O Cepheid \citep{ogleIV_cep_multi}. For recent review see \cite{pam14}.

The analysis of 1O Cepheids revealed another, most interesting group of double-periodic pulsators. In more than hundred 1O Cepheids additional small-amplitude variability, with period shorter than first overtone period, was detected. The period ratios fall in the $P/P_{\rm 1O}\in(0.6,\,0.65)$ range and cannot correspond to two radial modes \citep{wdrs}. 35 stars were reported in the Large Magellanic Cloud \citep[LMC,][]{mk09,ogle_cep_lmc}, 138 stars in the Small Magellanic Cloud \citep[SMC,][]{ogle_cep_smc} and 1 star was found in the Galactic disc \citep{pietruk}. One LMC star with additional variability pulsates simultaneously in the radial fundamental and first overtone modes \citep{mk09}. Two stars were also identified in the {\it Kepler}-{\it K2} photometry (Plachy et al., in prep.). In the Petersen diagram, i.e. in the plot of shorter-to-longer period ratio versus the longer period, these stars group in the three well separated sequences.

Interestingly, very similar form of pulsation is present in RR~Lyrae stars, see the most in-depth and extensive studies of the phenomenon by \cite{pamsm15}, \cite{netzel1,netzel3} and \cite{jurcsik_M3}. Additional variability is detected in first overtone pulsators (RRc) or in double-mode F+1O pulsators (RRd). Period ratios fall in the similar range, $P/P_{\rm 1O}\in(0.60,\,0.64)$. Three sequences are present in the Petersen diagram as well, although they are not that well separated as in the case of Cepheids \citep{netzel3}. Thanks to ultra-precise observations by space telescopes, {\it Kepler} and {\it CoRoT} \citep[e.g.][]{szabo_corot,pamsm15,molnar,kurtz}, and detailed analysis of ground-based observations \citep{netzel1,netzel3,jurcsik_M3} this form of pulsation is well studied in RR~Lyr stars. In particular, we know that in the frequency spectra of these stars signal (power excess) at subharmonic of the additional frequency is present. Signals associated with the additional variability are broad and non-coherent. In the time domain it corresponds to strong variability of amplitude and frequency on a time-scale of a few tens to hundred of days. The phenomenon must be common among RRc/RRd stars, as 14 out of 15 stars observed from space show this form of pulsation \citep[for a summary see][]{pamsm15}.

In contrast to RR~Lyr stars, 1O Cepheids were not extensively observed from space (see Sect.~\ref{ssec:rrlcomp}). Analysis of ground-based data, in particular of the largest sample of 138 of these interesting stars from SMC is missing. \cite{ogle_cep_smc} only reported the discovery of these stars and provided their periods and period ratios. In the present study we analyse the OGLE-III data for these interesting objects in detail. We do not search for additional objects, but focus on those in which we know that additional variability is present. Our goal is to study the properties of the variability in detail. In particular, we check for the presence of subharmonics of the additional signal, analyse the amplitude distribution and time-variation of the additional signals. These are necessary information for the models and theories to explain this peculiar and puzzling form of pulsation.

\section{Data analysis}\label{sec:methods}

We analyse OGLE-III $I$-band photometry for 138 stars listed in \cite{ogle_cep_smc}. All these stars were identified as 1O Cepheids with additional small amplitude variability, with period ratios in the $P/P_{\rm 1O}\in(0.6,\,0.65)$ range. We use standard consecutive pre-whitening technique. We identify the dominant frequencies with the help of discrete Fourier transform (FT). Next, we fit the data with the sine series of the following form:
\begin{align}
m(t)=m_0+\sum_k A_k\sin(2\pi\nu_kt+\phi_k)\,,\label{eq:ssum}
\end{align}
using the non-linear least-square fitting. The FT of residual data is inspected for the presence of additional signals, which are iteratively included in eq.~\eqref{eq:ssum}. Only resolved frequencies are included. We consider two peaks as well resolved if their separation is larger than $2/\Delta T$, where $\Delta T$ is data length. In the FT the signal is considered significant if signal-to-noise ratio ($S/N$) exceeds 4. The criterion is relaxed for signals at combination frequencies, provided that peak is present exactly (within frequency resolution) at the expected position (we accept $S/N>3.5$). Typically our solution consists of low-order ($3-6$) Fourier series describing the dominant variability associated with the first overtone ($k\fo$), sine term with the frequency of the additional variability of interest ($\fx$) and possibly with the combination frequency with the first overtone frequency (typically $\fo+\fx$). Additional significant signals we find, that do not fall in the $P/P_{\rm 1O}\in(0.6,\,0.65)$ range, are also included in eq.~\eqref{eq:ssum}.

During the analysis we reject the outliers ($4\sigma$ criterion) and remove slow trends through subtracting from the original data the low-order polynomials or splines. These functions are fitted to the residuals. Quite often, after prewhitening with the first overtone frequency and its harmonics, residual signal remains at the location unresolved with $k\fo$. Typically it corresponds to the long-term variation of the first overtone phase (period change). This signal may be significant which increases the noise level in the FT and consequently may hide the additional variabilities. In such case we get rid of the non-stationary first overtone variation with the help of time-dependent prewhitening technique, described and applied to the {\it Kepler} data by \cite{pamsm15}. Application to the ground-based OGLE data is described in more detail in \cite{netzel1}.

Strong daily aliases and 1-yr aliases are inherent to ground-based OGLE observations of the SMC. As the signals we search for are relatively weak, the alias-related ambiguities can happen. In some stars, after prewhitening with the first overtone frequency, we detect a few significant peaks of similar height which are mutual daily aliases. If period corresponding to one of them falls in the $P/P_{\rm 1O}\in(0.6,\,0.65)$ range, then this peak is accepted as a true signal, even if it is not the highest peak. All such cases are reported explicitly in the study.

\section{Results}\label{sec:results}

\subsection{Overview}\label{ssec:overview}

\begin{table*}
\caption{Properties of 1O Cepheids with additional variability. The consecutive columns contain: star's id, first overtone period, $\Po$, period of the additional variability, $\Px$, period ratio, $\pxpo$, amplitude of the first overtone, $A_{\rm 1O}$, and amplitude ratio, $A_{\rm x}/A_{\rm 1O}$, and remarks: `al' -- daily alias of signal at $\fx$ is higher; `nsx' -- complex appearance of the signal at $\fx$; `nsO' -- non-stationary first overtone; `cf' -- combination frequency of $\fx$ and $\fo$ detected; `sh' -- power excess at subharmonic frequency (centred at $1/2\fx$) detected; `ap' -- additional periodicity detected; `tdp' -- time-dependent analysis was conducted; `?' -- weak detection ($S/N$ given in the parenthesis). Full Table is in the Appendix~\ref{app:table} (Tab.~\ref{tab:atab}).}
\label{tab:tab}
\begin{tabular}{lrrrrrr}
star & $\Po$\thinspace (d) & $\Px$\thinspace (d) & $\pxpo$ & $A_{\rm 1O}$\thinspace (mag) & $A_{\rm x}/A_{\rm 1O}$ & remarks \\
\hline
OGLE-SMC-CEP-0056  & 0.9860208(7) & 0.60373(1) & 0.6123 & 0.1689 & 0.024 & ? ($S/N=3.77$) \\
OGLE-SMC-CEP-0212  & 1.741010(4)  & 1.08766(4) & 0.6247 & 0.0997 & 0.036 & sh, nsx \\
OGLE-SMC-CEP-0251  & 1.796802(1)  & 1.12279(2) & 0.6249 & 0.1399 & 0.029 & sh, nsx \\
OGLE-SMC-CEP-0280  & 1.675191(1)  & 1.04344(2) & 0.6229 & 0.1377 & 0.026 & nsO, ap \\
OGLE-SMC-CEP-0281  & 1.2662457(7) & 0.774075(9)& 0.6113 & 0.1263 & 0.033 & al, nsx \\
OGLE-SMC-CEP-0307  & 0.9734743(7) & 0.59718(1) & 0.6134 & 0.1922 & 0.019 & nsO, ap \\
OGLE-SMC-CEP-0447  & 1.2651448(8) & 0.77624(1) & 0.6136 & 0.1300 & 0.024 & nsO, nsx \\
\ldots & & & & & & \\
\hline
\end{tabular}
\end{table*}

\begin{figure}
\centering
\resizebox{\hsize}{!}{\includegraphics{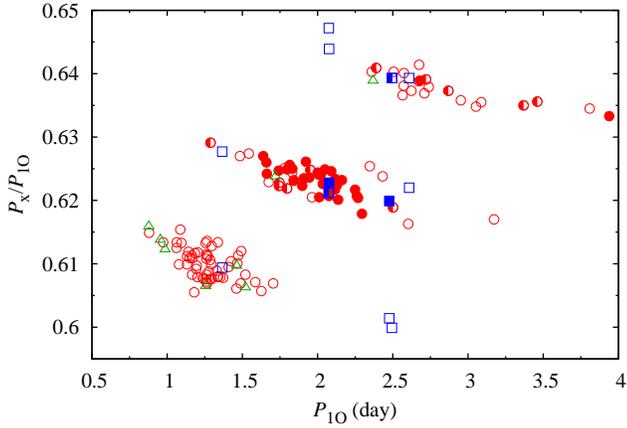}}
\caption{Petersen diagram for 138 analysed Cepheids. Stars in which two periodicities were detected, corresponding to two sequences in the diagram, are marked with squares (two squares per star). Stars with weak detection of the additional periodicity are marked with open triangles. Filled symbols correspond to stars in which power excess centred at subharmonic frequency, $\fx/2$, was detected (half-filled symbols are used to indicate weak detections).}
\label{fig:pet}
\end{figure}

Results of our analysis are collected in Tab.~\ref{tab:atab} in the Appendix. For a reference a section of the Table is presented in Tab.~\ref{tab:tab}. In consecutive columns there are: star's id, first overtone period, $\Po$, period of the additional variability in the $P/\Po\in(0.6,\,0.65)$ range, $\Px$, period ratio, $\pxpo$, amplitude of the first overtone mode, amplitude ratio, $A_{\rm x}/A_{\rm 1O}$, and remarks. The resulting Petersen diagram is plotted in Fig.~\ref{fig:pet}. 

\begin{figure}
\centering
\resizebox{\hsize}{!}{\includegraphics{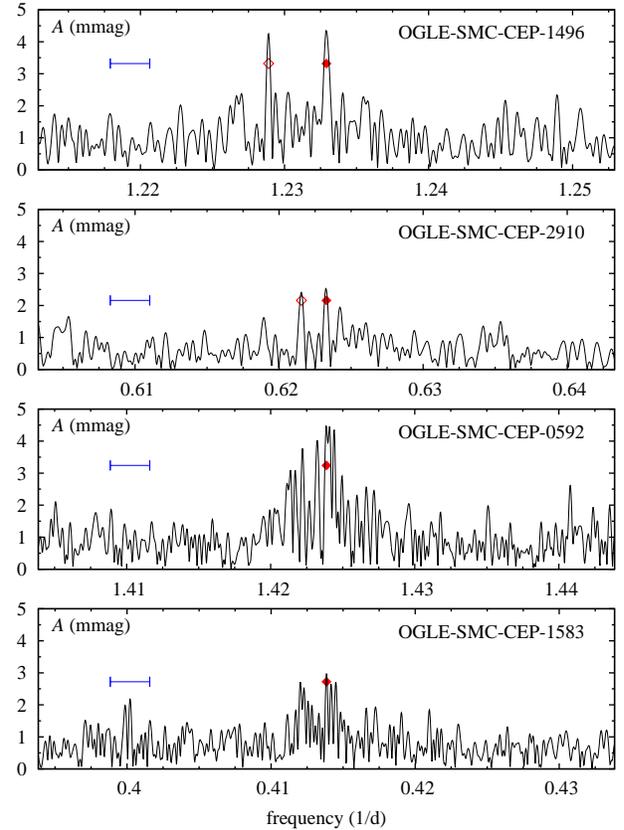}}
\caption{Illustration of complex structures detected at $\fx$. In the two upper panels, two well separated peaks of similar height are present, while in the two lower panels, clusters of peaks are present. The extent of the horizontal bar corresponds to the separation expected for 1-yr alias. Filled diamonds mark the location of peaks included in Tab.~\ref{tab:tab}. In the two upper panels, open diamonds indicate the location of peaks adopted by Soszy\'nski et al. (2010) (see Appendix, Sect.~\ref{ssec:igor}).}
\label{fig:ilu}
\end{figure}

In the frequency spectrum, the additional variability rarely appears as a single and coherent peak. Typically more complex structures are present; examples are illustrated in Fig.~\ref{fig:ilu}. Sometimes two dominant close peaks are detected, as illustrated in the top two panels of Fig.~\ref{fig:ilu}. In other cases the signal appears as a complex cluster of peaks, as illustrated in the two lower panels of Fig.~\ref{fig:ilu}. In our analysis we pick the highest peak in the cluster, or in a group of close peaks (marked with filled diamonds in Fig.~\ref{fig:ilu}), and include its frequency in eq.~\eqref{eq:ssum}. Its properties, period and amplitude, $\Px$ and $A_{\rm x}$, are then given in Tab.~\ref{tab:tab}. After prewhitening, residual, unresolved power is often detected. Such appearance of additional variability indicates that it is strongly non-stationary, with variable phase and/or amplitude (see Section~\ref{ssec:tv}). All stars in which more complex structures are detected at $\fx$ (two close peaks, clusters of peaks, residual power after prewhitening) are marked with `nsx' in the remarks column of Tab.~\ref{tab:tab}. These structures indicate that additional variability is non-stationary, see Sect.~\ref{ssec:tv}. The different appearance of the signal may result from different time-scales of the variability and different structure of the data (different length of the available data). There are 80 stars marked with `nsx' which is 58\thinspace per cent of the analysed sample. The complex structures at $\fx$ are common.

First overtone is often non-stationary as well, which appears as strong unresolved power at its frequency, after the prewhitening. These stars are marked with `nsO' in Tab.~\ref{tab:tab}. There are 56 such stars which constitutes nearly 41\thinspace per cent of the analysed sample. In some cases the unresolved power at $\fo$ dominates the frequency spectrum and significantly increases the noise level in the Fourier transform, which may hide the additional significant peaks. In all such cases we conducted time-dependent prewhitening to get rid of the unwanted signal. If time-dependent prewhitening was crucial for the detection of additional variability of interest, or significantly improved the $S/N$ of the interesting peak at $\fx$, then `tdp' is included in the remarks column of Tab.~\ref{tab:tab}. In these cases the frequency and amplitude of the additional variability are determined independently of the determination of amplitude and frequency of the first overtone, from the dataset with first overtone filtered out. An inherent part of the time-dependent prewhitening is time-dependent Fourier analysis, which shows how the amplitude and phase of the first overtone change in time. In the majority of cases we observed a pronounced phase change, while amplitude changes were insignificant. No firm case of Blazhko-like modulation was found, although in some cases variation of first overtone phase seemed periodic.

Fast period changes, on  time-scale shorter than expected from evolutionary calculations are common in 1O Cepheids. Exemplary O-C diagrams may be found e.g. in studies by \cite{berd1} or \cite{berd2}. A detailed study of OGLE and MACH data for LMC Cepheids was conducted by \cite{poleskiPC} who detected period changes in 41 per cent of 1O LMC Cepheids and in 18 per cent of fundamental mode pulsators. The analysis of period changes of the first overtone is beyond the scope of the present analysis, however; dedicated study is planned. 

In 25 stars we find peaks at combination frequency, $\fo+\fx$, and in one star we find peak at $\fx-\fo$ (OGLE-SMC-CEP-0797). These stars are marked with `cf' in the last column of Tab.~\ref{tab:tab}. 

Stars in which signal at a daily alias of $\fx$ is higher are marked with `al'. As described in the previous section, we select the lower alias as a true signal if it falls well within one of the three sequences in the Petersen diagram.  

In 17 stars marked with `ap' in Tab.~\ref{tab:tab} we detect additional significant periodicity that does not fall into the $P/\Po\in(0.60,\,0.65)$ range and cannot be interpreted as due to other radial mode. In a few stars the additional peaks appear relatively close to the radial first overtone frequency. We note that similar detections were reported by \cite{mk09}, who argue that these signals may be intrinsic to the stars and correspond to non-radial pulsation. In no case we detect combination frequency with the first overtone, however. In principle these periodicities may result from blending. We postpone the discussion of these additional signals till the analysis of full sample of SMC Cepheids, which will allow to draw some statistically meaningful conclusions concerning their nature.  

In stars plotted with triangle in Fig.~\ref{fig:pet} the detection of additional variability is weak. These stars are marked with `?' and $S/N$ value is given in the last column of Tab.~\ref{tab:tab}. In some cases the additional signal appeared only after the time-dependent prewhitening. There are eight such cases and we discuss them in more detail in the Appendix, in Sect.~\ref{ssec:igor}, which also contains detailed comparison of our results with those reported in \cite{ogle_cep_smc}. The period ratios for these stars fall well within the three sequences in the Petersen diagram. Despite the weak detection, we consider these stars as double-periodic in the following. 

\begin{figure}
\centering
\resizebox{\hsize}{!}{\includegraphics{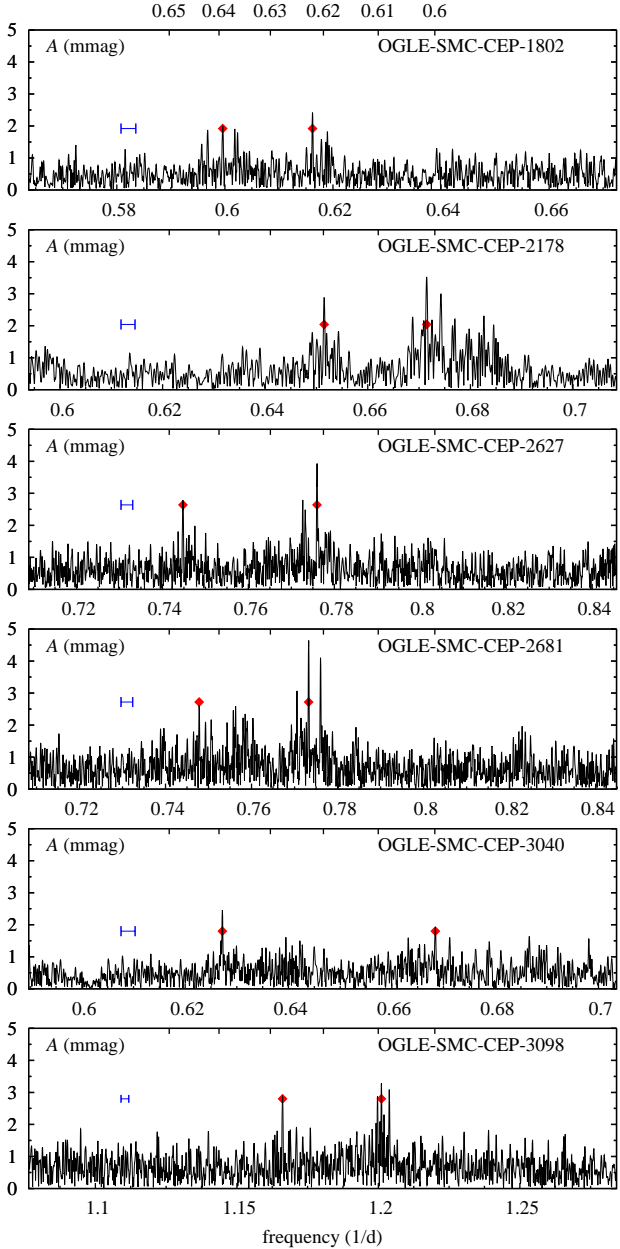}}
\caption{Frequency spectra for six stars in which two significant peaks, corresponding to two sequences in the Petersen diagram, were detected. These peaks are marked with filled diamonds placed at the $S/N=4.0$ level. The extent of the horizontal bar, plotted in each panel, corresponds to separation expected for 1-yr aliases. Period-ratio scale, $P/\Po$, is plotted at the top of each panel, with numerical labels plotted in the top-most panel.}
\label{fig:2seq}
\end{figure}

In six stars, two well separated and significant peaks were detected in the frequency range of interest -- the corresponding period ratios fall within two separate sequences in Fig.~\ref{fig:pet}. In the figure these stars are marked with squares, two for each of six stars. Their frequency spectra are plotted in Fig.~\ref{fig:2seq}. Typically, signal corresponding to one of the sequences is dominant, while the detection of peak corresponding to the other sequence is rather weak (but always with $S/N>4.0$). In Tab.~\ref{tab:tab}, two rows are present for these stars, with characteristics of the highest peaks falling within one of the sequences. All other stars, in which we detect a significant peak corresponding to only one sequence, are marked with circles in Fig.~\ref{fig:pet}.

Finally, in many stars we detect significant power centred at $1/2\fx$, i.e. at subharmonic frequency. These stars are marked with `sh' in the remarks column of Tab.~\ref{tab:tab}, printed in {\it italics} if the detection is weak. In the Petersen diagram in Fig.~\ref{fig:pet}, stars with firm detection of the power excess at subharmonic frequency are marked with filled symbols. In the case of weak detection half-filled symbol is plotted. These signals will be discussed in detail in Section~\ref{ssec:sh}.

\subsection{The Petersen diagram and amplitude distributions}\label{ssec:amps}

Three well separated and slanted sequences are present in the Petersen diagram (Fig.~\ref{fig:pet}). 64 stars fall within the bottom sequence, 54 stars fall within the middle sequence and 26 stars fall within the top sequence. The numbers do not add up to 138, as six stars fall within two sequences simultaneously. Within each sequence period ratio drops with the increasing pulsation period. Stars forming the bottom sequence have, on average, shorter pulsation periods, while stars forming the top sequence have, on average, longer pulsation periods. 

\begin{figure}
\centering
\resizebox{\hsize}{!}{\includegraphics{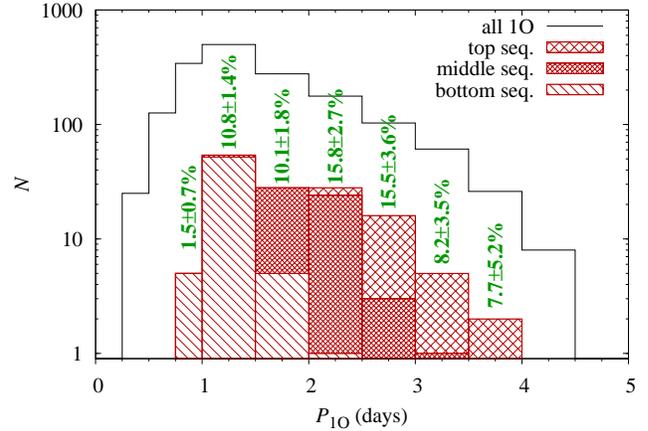}}
\caption{Period distribution for all 1O SMC OGLE-III Cepheids (solid black line) and for stars with the additional variability (hatched area). Contributions from the three sequences in the Petersen diagram are marked with different patterns, as indicated in the key. Incidence rates are also provided.}
\label{fig:histoX}
\end{figure}

The number of stars in the top sequence is significantly smaller than in the middle and bottom sequences. On the other hand, long-period first overtone Cepheids are not as numerous as short-period overtone pulsators. In Fig.~\ref{fig:histoX}, we study the period distribution for all 1644 first overtone SMC Cepheids from OGLE-III collection (solid black line) and for stars with additional variability (hatched area; three different patterns show the contributions from the three sequences). Stars were counted in 0.5\thinspace d-wide bins, except $\Po<1$\thinspace d, where we used smaller, 0.25\thinspace d-wide bins. This is because of sharp increase of Cepheid number as one moves from $0.25$-$0.5$\thinspace d bin  through $0.5$-$0.75$\thinspace d bin to $0.75$-$1.0$\thinspace d bin. Within each bin the incidence rate of stars with additional variability is given with statistical errors calculated assuming that the population follows a Poisson distribution \citep[e.g.][]{alcock}. 

It is well visible that the discussed form of pulsation is not present in the shortest period 1O Cepheids with $\Po\!<\!0.75{\rm d}$, despite of 151 stars falling into this period range. For $0.75{\rm d}\!<\!\Po\!<\!1.0{\rm d}$ there are 342 1O Cepheids and only in five of them the additional variability was found. Incidence rate is very low ($1.5\pm0.7$\thinspace per cent) as compared to the next longer-period bin ($10.8\pm1.4$\thinspace per cent). This is most likely due to selection effect. Shortest period overtone Cepheids are least luminous (because of the $P-L$ relation, see Fig.~\ref{fig:cwa_basics}), consequently one may expect higher noise level in the Fourier transform, which may hinder the detection of low-amplitude variability. This is discussed in more detail in Sect.~\ref{ssec:selection}. For $\Po\!>\!1{\rm d}$ the discussed form of pulsation is quite frequent. For $1{\rm d}\!\leq\!\Po\!<\!2{\rm d}$ the incidence rate is $\approx\!10.5$\thinspace per cent, for $2{\rm d}\!\leq\!\Po\!<\!3{\rm d}$ it is $\approx\!15.5$\thinspace per cent and for $3{\rm d}\!\leq\!\Po\!<\!4{\rm d}$ it is $\approx\!8$\thinspace per cent. Taking into account the statistical errors, these numbers are not that different. We conclude that the top sequence is the least numerous of the three, mostly because there are fewer long-period 1O Cepheids than short-period ones and also because the incidence rate may be slightly lower for longer periods.

\begin{figure}
\centering
\resizebox{\hsize}{!}{\includegraphics{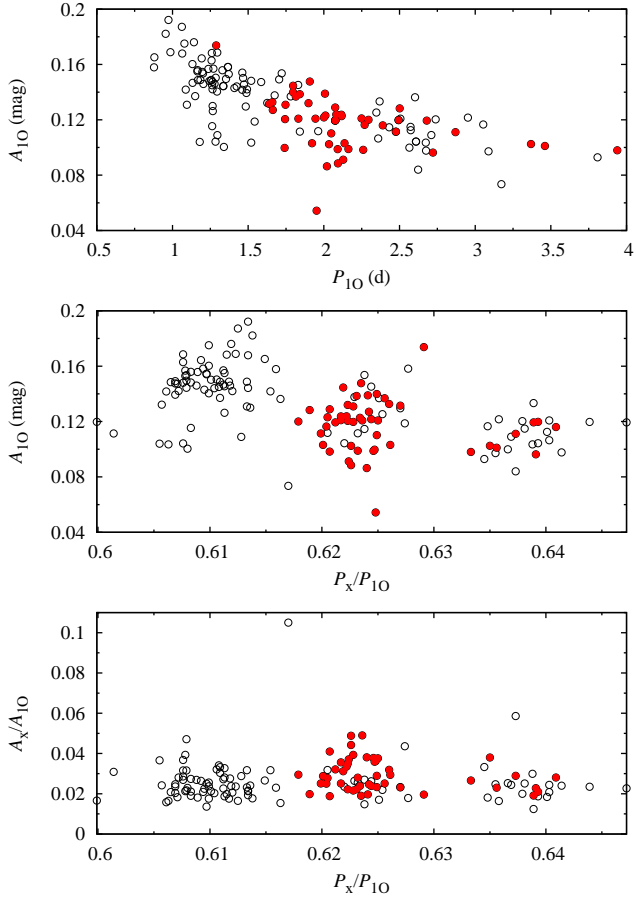}}
\caption{First overtone amplitude as a function of $\Po$ (top panel) and $\pxpo$ (middle panel). In the bottom panel we show $A_{\rm x}/A_{\rm 1O}$ as a function of $\pxpo$. Stars in which power excess at subharmonic, $1/2\fx$, was detected are plotted with filled symbols.}
\label{fig:amps}
\end{figure}

\begin{figure}
\centering
\resizebox{\hsize}{!}{\includegraphics{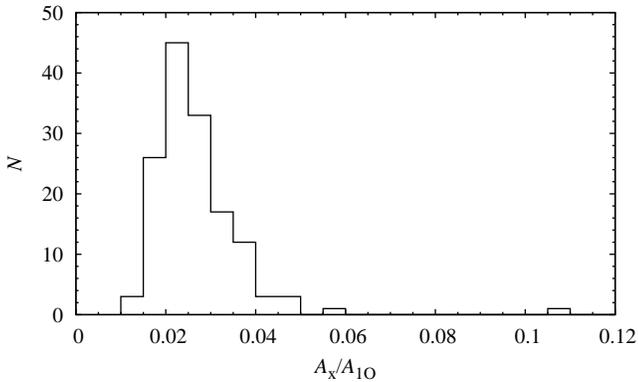}}
\caption{Histogram of  $A_{\rm x}/A_{\rm 1O}$ values.}
\label{fig:hia}
\end{figure}

The additional variability is always weak as compared to radial first overtone. Tab.~\ref{tab:tab} provides the amplitude of the first overtone, $A_{\rm 1O}$, and amplitude ratio, $A_{\rm x}/A_{\rm1O}$. The top panel of Fig.~\ref{fig:amps} shows the amplitude of the first overtone as a function of the first overtone period. The amplitude drops with increasing pulsation period. The highest (Fourier) amplitude is slightly below $0.2$\thinspace mag, the lowest is around $0.06$\thinspace mag. The most typical values fall in the $0.10-0.16$\thinspace mag range. In the middle panel of Fig.~\ref{fig:amps} we plot the amplitude of the first overtone as a function of period ratio, $\pxpo$. The trace of the three sequences present in the Petersen diagram is well visible. It is clear that amplitudes are the highest in stars forming the bottom sequence (as these stars have shorter first overtone periods, on average) and the lowest in the stars forming the top sequence (stars with longer first overtone periods). The bottom panel of Fig.~\ref{fig:amps} shows the amplitude ratio, $A_{\rm x}/A_{\rm1O}$, as a function of period ratio. There is no significant difference between the stars corresponding to the three sequences. In stars that are plotted with filled symbols in Fig.~\ref{fig:amps}, a power excess at subharmonic, $1/2\fx$, is detected. This will be discussed in more detail in Sect.~\ref{ssec:sh}. Here we just note that for stars of the middle and top sequences (only there power excess at subharmonic was detected) there is no difference in first overtone amplitude between stars with, and stars without power excess at subharmonic. The histogram of amplitude ratios, $A_{\rm x}/A_{\rm 1O}$, for all the stars is presented in Fig.~\ref{fig:hia}. The distribution peaks at $A_{\rm x}/A_{\rm 1O}\approx 1.75-2.25$\thinspace per cent. Typical amplitudes of the additional variability are in the 2--5\thinspace mmag range (see also Fig.~\ref{fig:sh_amps}). It explains why the additional variability was discovered only recently -- high-quality observations are necessary to detect such low-amplitude variability. So far, the signal was detected mostly in the excellent OGLE data. Additional variability was also detected in two 1O Cepheids observed with {\it K2} (Plachy et al., in prep.).

\subsection{Comparison with other first overtone Cepheids without additional variability}\label{ssec:cwa}

\begin{figure}
\centering
\resizebox{\hsize}{!}{\includegraphics{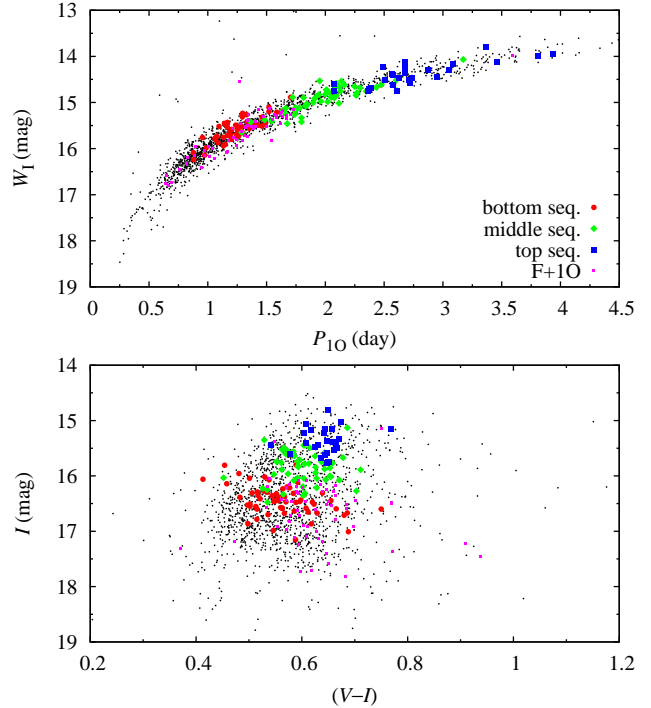}}
\caption{Period-luminosity (top panel) and colour-magnitude (bottom panel) diagrams for all 1O Cepheids from the SMC OGLE-III collection (small dots) and for Cepheids with additional variability analysed in this paper (larger symbols; red circles correspond to the bottom sequence in the Petersen diagram, green diamonds to the middle sequence and blue squares to the top sequence). In addition, double-mode, F+1O Cepheids are plotted with small squares.}
\label{fig:cwa_basics}
\end{figure}

In this Section, we check whether Cepheids with additional variability differ significantly from other first overtone SMC Cepheids, that are single-periodic. In Fig.~\ref{fig:cwa_basics}, we plot the location of analysed stars in the period-luminosity (top panel) and in the colour-magnitude (bottom panel) diagrams. For the former diagram we use reddening free Wesenheit index as the luminosity indicator. In the plots, all 1O Cepheids are plotted with small black dots, while stars with additional variability are plotted with larger symbols; point shape and point colour code the star's location in the Petersen diagram (red circles -- bottom sequence, green diamonds -- middle sequence and blue squares -- top sequence). Stars in which variability corresponding to two sequences was detected, are plotted only once, as members of the sequence for which corresponding amplitude is higher. For this plot, periods and intensity mean $I$- and $V$-band magnitudes were taken directly from the OGLE-III ftp archive \citep{ogle_cep_smc}.

\begin{figure}
\centering
\resizebox{\hsize}{!}{\includegraphics{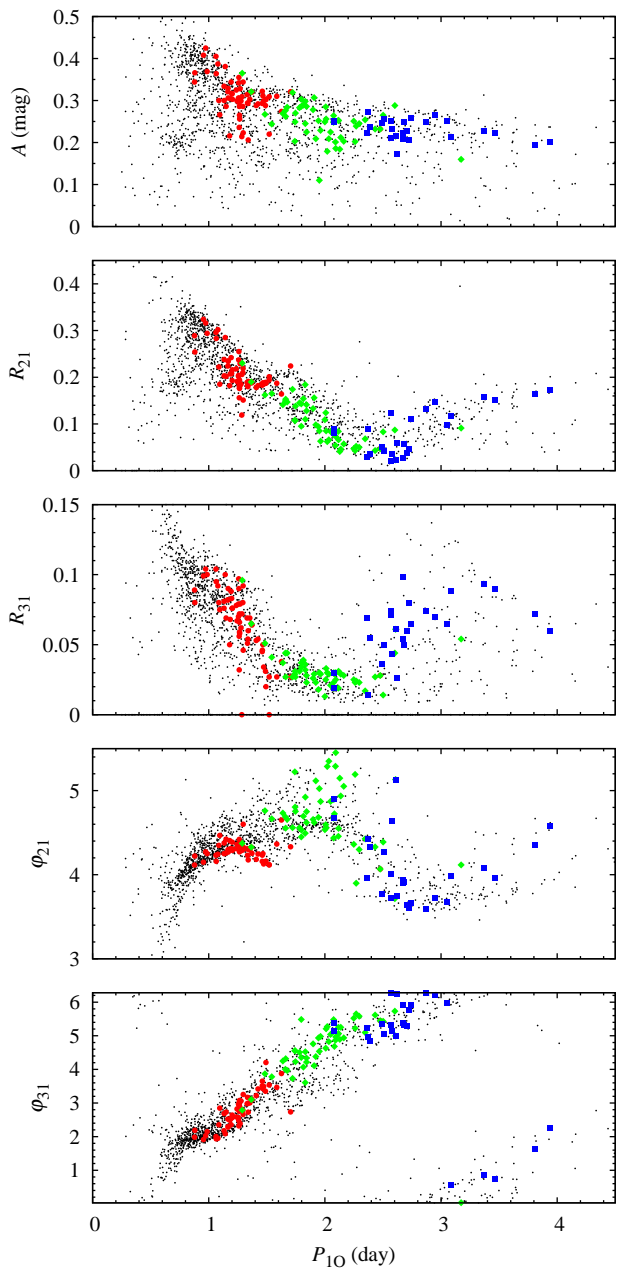}}
\caption{Fourier decomposition parameters versus the first overtone period for all 1O Cepheids from the SMC OGLE-III collection (small dots) and for Cepheids with additional variability analysed in this paper (larger symbols, as in Fig.~\ref{fig:cwa_basics}).}
\label{fig:cwa_fp}
\end{figure}

Except the lack of additional variability in the shortest period stars ($\Po<0.8$\thinspace d), which we already noted in the previous Section, we see no significant differences in the distribution of stars with and without the additional variability in the two diagrams presented in Fig.~\ref{fig:cwa_basics}. They cover similar colour and Wesenheit index ranges. Separation of stars, members of the three sequences in the Petersen diagram, is clear and pronounced. As period-luminosity diagram indicates, it results from different first overtone periods characteristic for the three groups. Members of the bottom sequence have shorter periods and consequently are least luminous, while members of the top sequence have longer periods and are thus most luminous. For stars of each sequence the covered colour range is similar as for single-periodic stars of similar luminosity. In Fig.~\ref{fig:cwa_basics}, we also plotted double-mode F+1O Cepheids (small squares). These stars cover similar luminosity range as stars of the bottom sequence, but are shifted, on average, towards higher colour values. 

In Fig.~\ref{fig:cwa_fp}, we compare the light curve shapes with the help of the Fourier decomposition parameters \citep{sl81}. Symbols used in the panels are exactly the same as in Fig.~\ref{fig:cwa_basics}. In the consecutive panels, from top to bottom, we plot: peak-to-peak amplitude, $R_{21}$, $R_{31}$, $\varphi_{21}$ and $\varphi_{31}$, all as a function of first overtone period. A lack of short period Cepheids with additional variability is again apparent. Also, additional variability is not detected in stars with low first overtone amplitude, $A\lesssim 0.2$\thinspace mag (and consequently in stars with lower $R_{21}$ and $R_{31}$), which is a selection effect. First, these stars are not as numerous as higher amplitude Cepheids. Second, with typical amplitude of the additional variability corresponding to $\sim\!2-4$\thinspace per cent of the first overtone amplitude, possible signals are also of lower amplitude and likely hidden in the noise. Otherwise, it seems that Cepheids with additional variability follow the same progressions as Cepheids without additional variability. The only exception seems the behaviour of $\varphi_{21}$ in the narrow period range of $1.4-1.6$\thinspace d. In this period bin, 1O Cepheids cover the $4\lesssim\varphi_{21}\lesssim 5$ range, but stars with additional variability and of the bottom sequence (red circles) prefer the low values, $\varphi_{21}\lesssim 4.3$.

We conclude that there is no significant difference between 1O Cepheids with and without the additional variability, with regards to their location on the period-luminosity and colour-magnitude diagrams, and light curve shapes (with the possible exception of $\varphi_{21}$ in relatively narrow period range).

\subsection{Subharmonics}\label{ssec:sh}

In many stars we detect significant signal centred at $1/2\fx$, i.e. at subharmonic of $\fx$. Typically the signal detected at $1/2\fx$ has a complex form -- broad cluster of peaks, centred at $1/2\fx$ is detected. Stars in which such power excess was detected are marked with `sh' in Tab.~\ref{tab:tab}. The weak detection is marked with `{\it sh}'. What we regard as `weak' is somewhat subjective. In general, if $3.5\!<\!S/N\!<\!4.0$ for the highest peak at around $1/2\fx$, or the power excess was evident only after the time-dependent prewhitening, the star is marked as a weak detection. In all cases however, the power excess was clear. Altogether, it was detected in 48 stars, of which 14 are marked as weak detections. This constitutes 35\thinspace per cent of the analysed sample or, if weak detections are excluded, 25\thinspace per cent. Detailed characterization of frequency spectra of stars with power excess detected at subharmonic frequency is collected in Tab.~\ref{tab:sh}. Power excess is characterized by the frequency and amplitude of the highest peak detected at around $1/2\fx$, $\fsh$ and $A_{\rm sh}$, respectively. Table contains: star's id, period ratio, $\pxpo$, frequency of the additional variability, $\fx$, frequency of the highest peak detected around $1/2\fx$, $\fsh$, amplitude of the additional variability, $A_{\rm x}$, and amplitude ratio, $A_{\rm sh}/A_{\rm x}$, approximate $S/N$ for the peak at $\fsh$ and remarks: `weak' -- weak detection, `broad' -- broad power excess; `tdp' -- time-dependent prewhitening of all signals except $\fsh$ conducted.

\begin{figure*}
\centering
\noindent\resizebox{0.33\hsize}{!}{\includegraphics{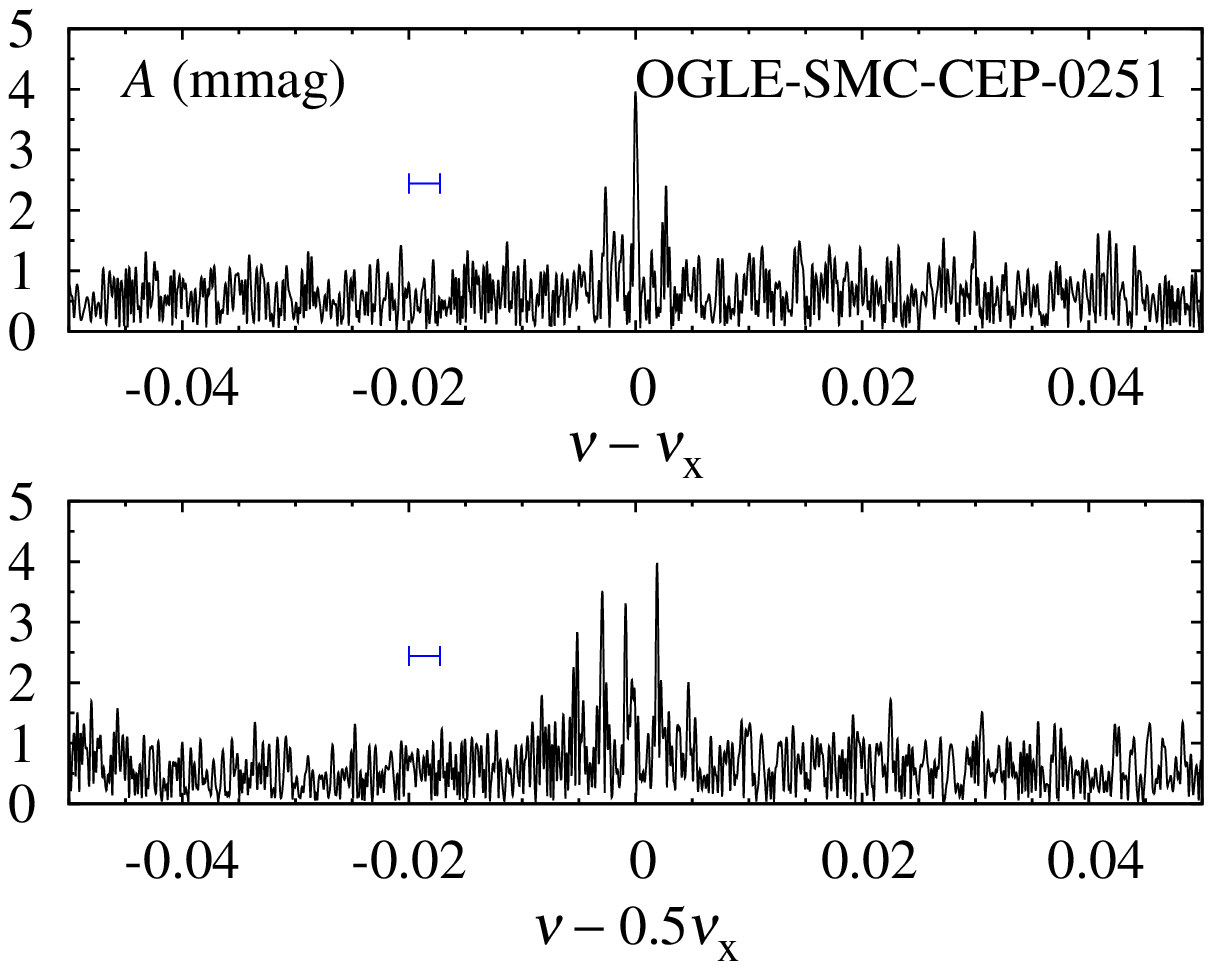}}
\resizebox{0.33\hsize}{!}{\includegraphics{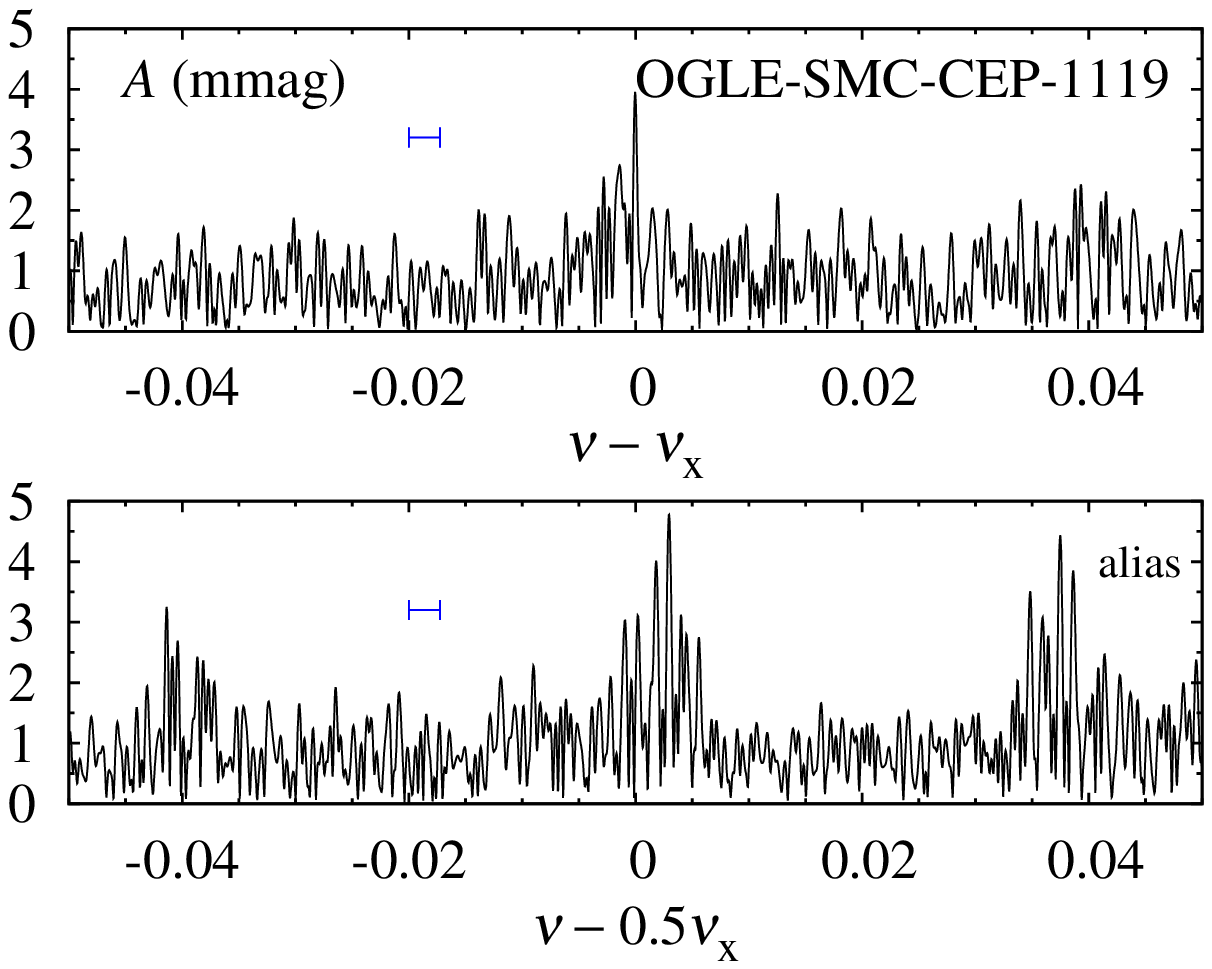}}
\resizebox{0.33\hsize}{!}{\includegraphics{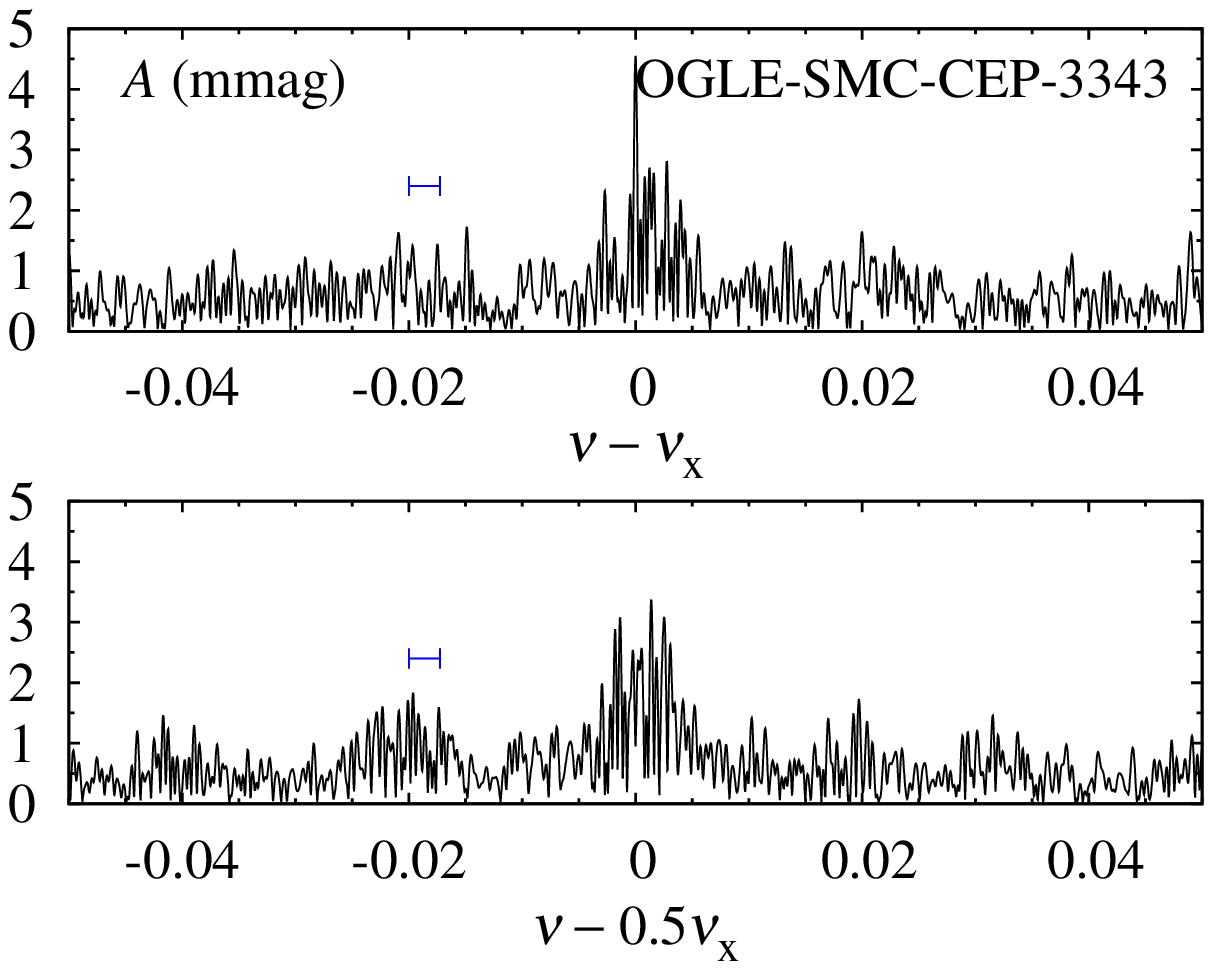}}\\
\noindent\resizebox{0.33\hsize}{!}{\includegraphics{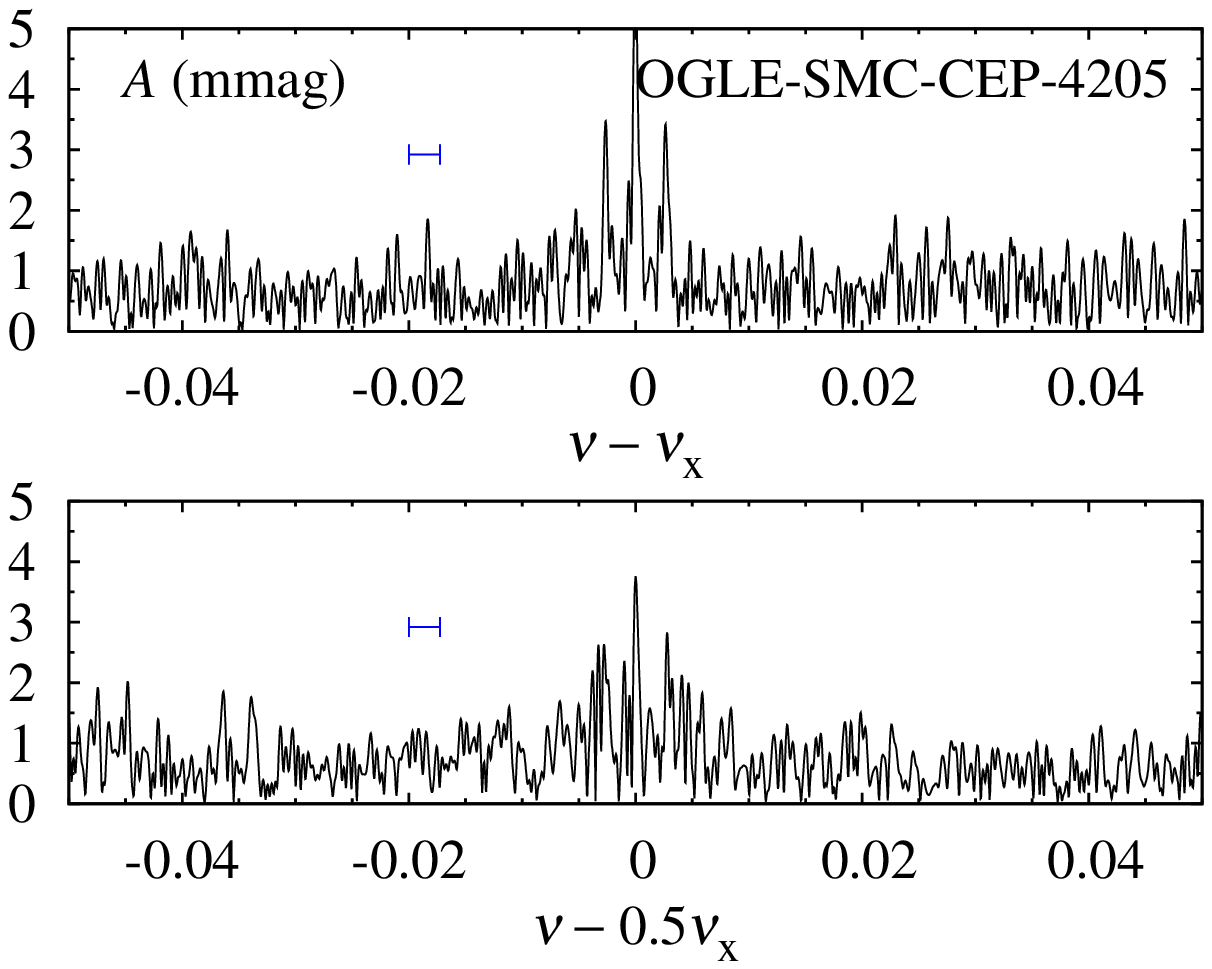}}
\resizebox{0.33\hsize}{!}{\includegraphics{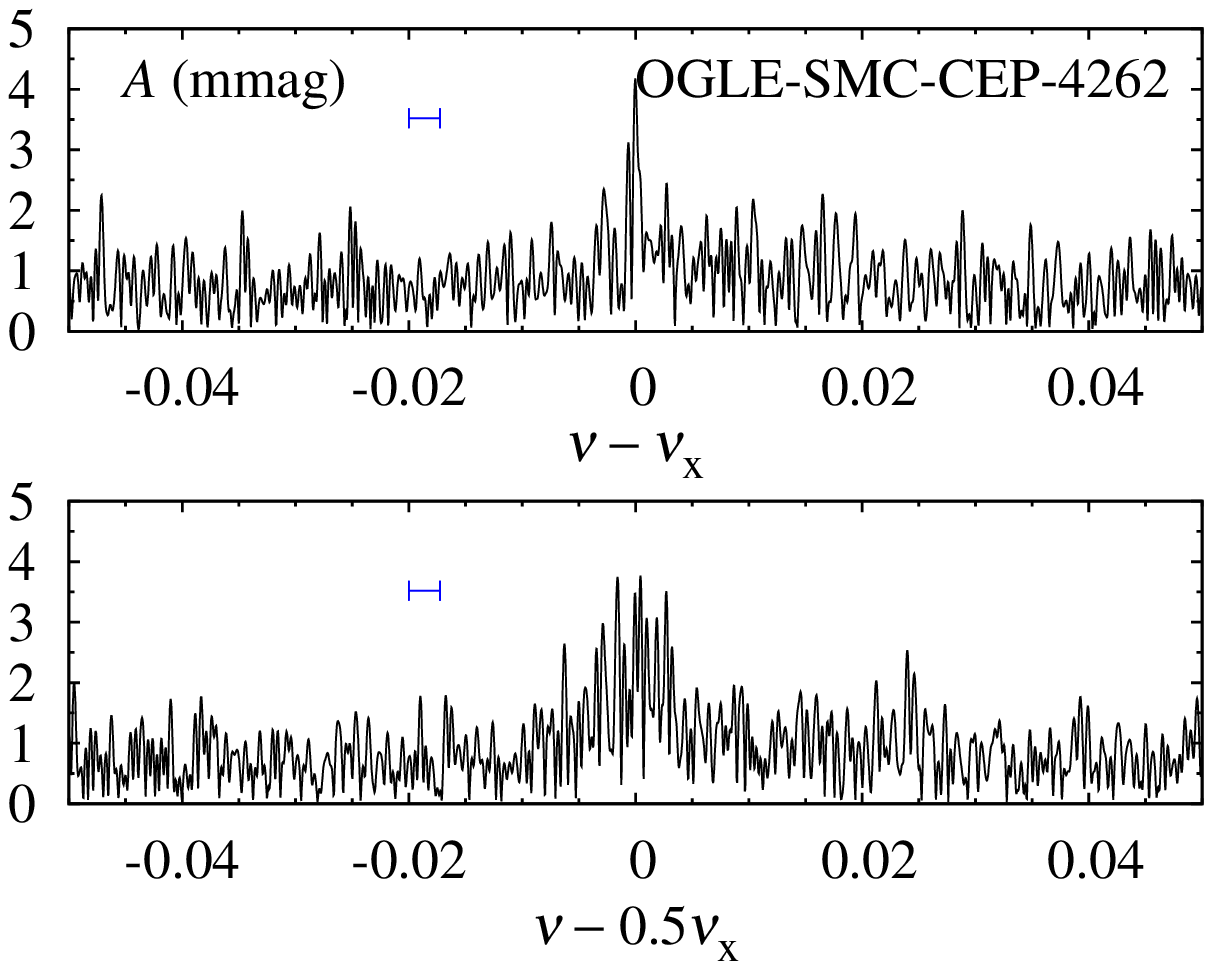}}
\resizebox{0.33\hsize}{!}{\includegraphics{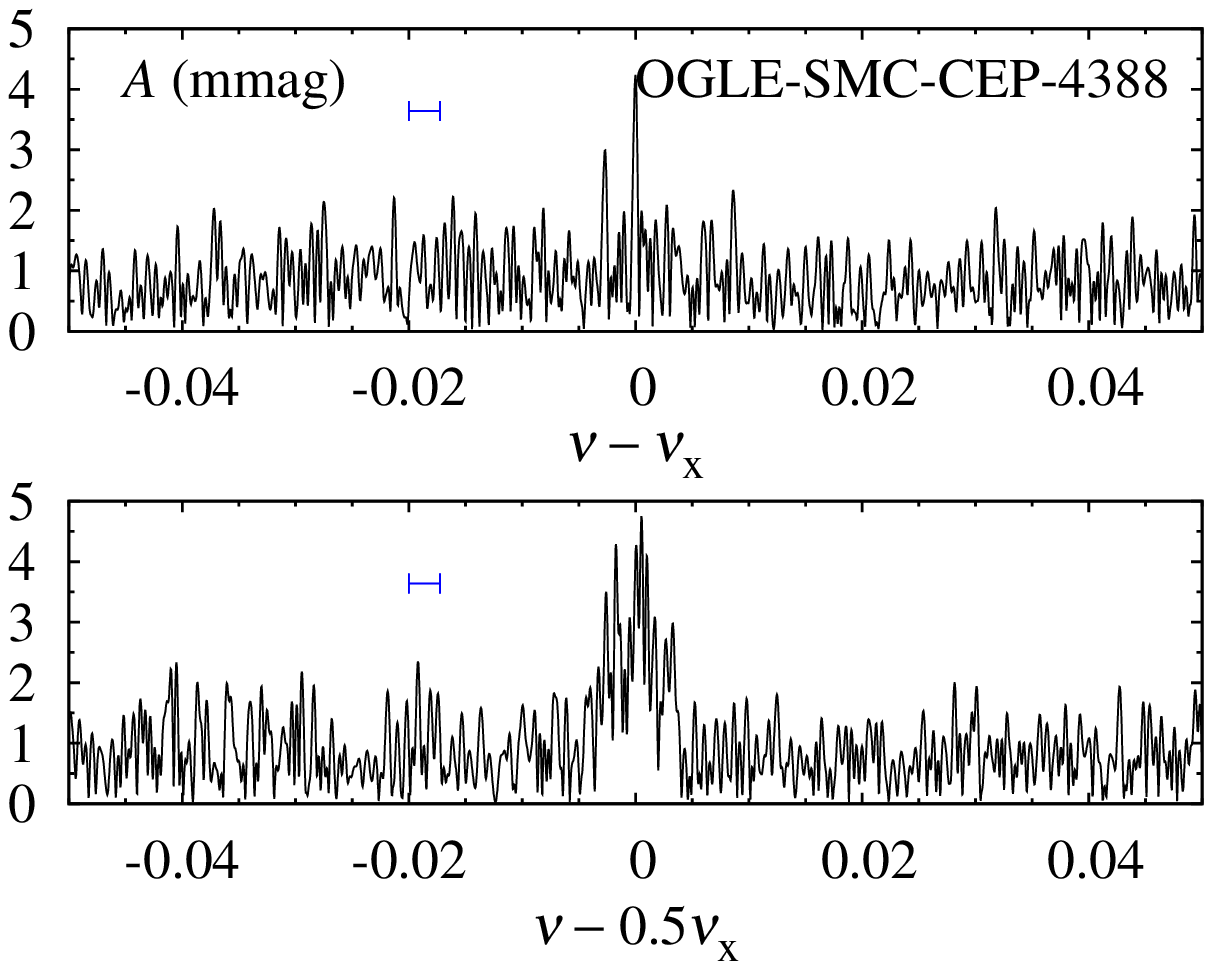}}
\caption{Frequency spectra for selected stars with additional variability present at $\fx$ and simultaneously with significant power excess centred at $1/2\fx$. In the top panel a section of frequency spectrum centred at $\fx$ is plotted. In the bottom panel a section of frequency spectrum centred at $\fx/2$ is plotted. The frequency range is the same in two panels. The extent of the horizontal bar plotted in each panel (at $S/N=4.0$) corresponds to separation expected for 1-yr aliases.}
\label{fig:sh_nice}
\end{figure*}

\begin{figure*}
\centering
\noindent\resizebox{0.33\hsize}{!}{\includegraphics{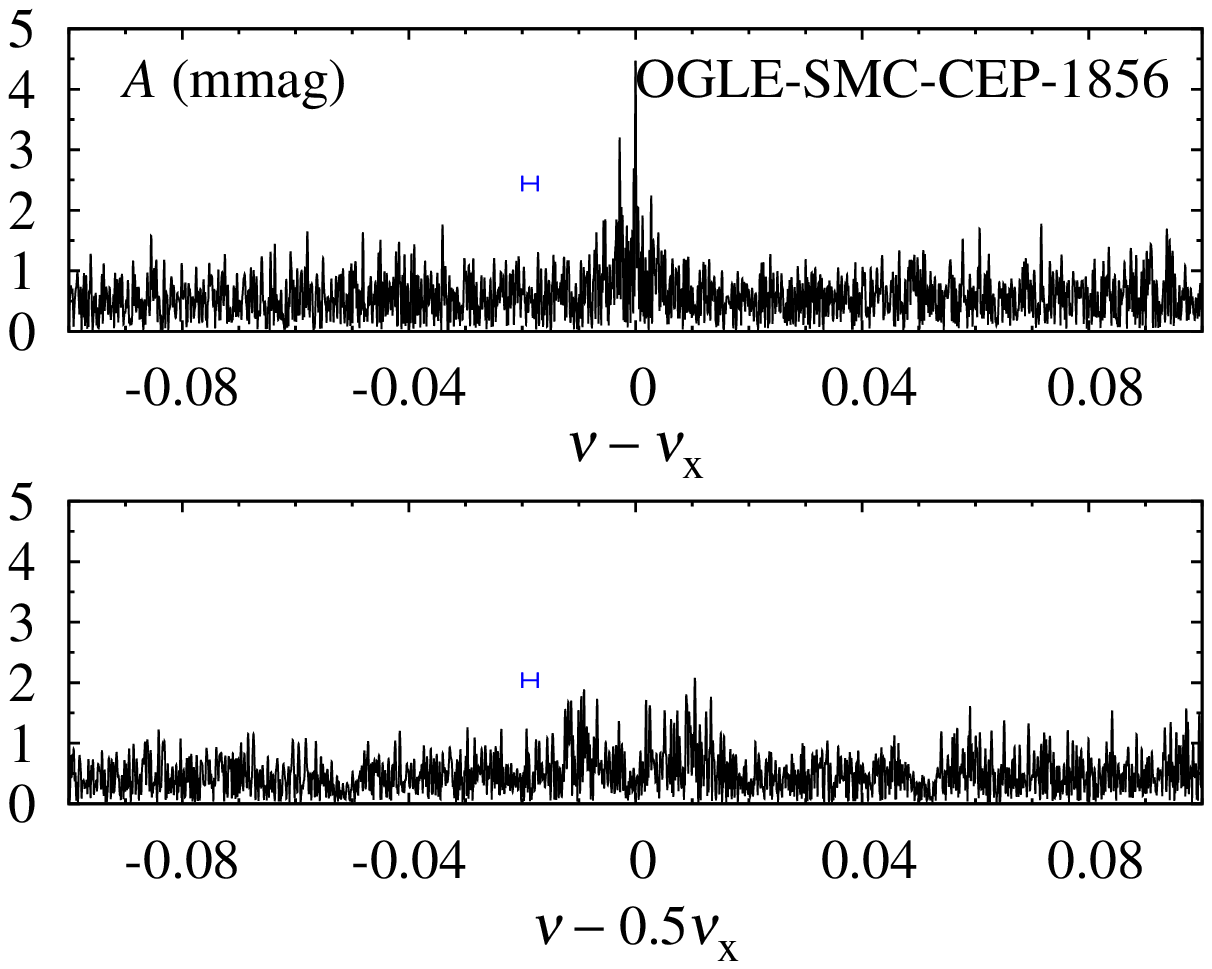}}
\resizebox{0.33\hsize}{!}{\includegraphics{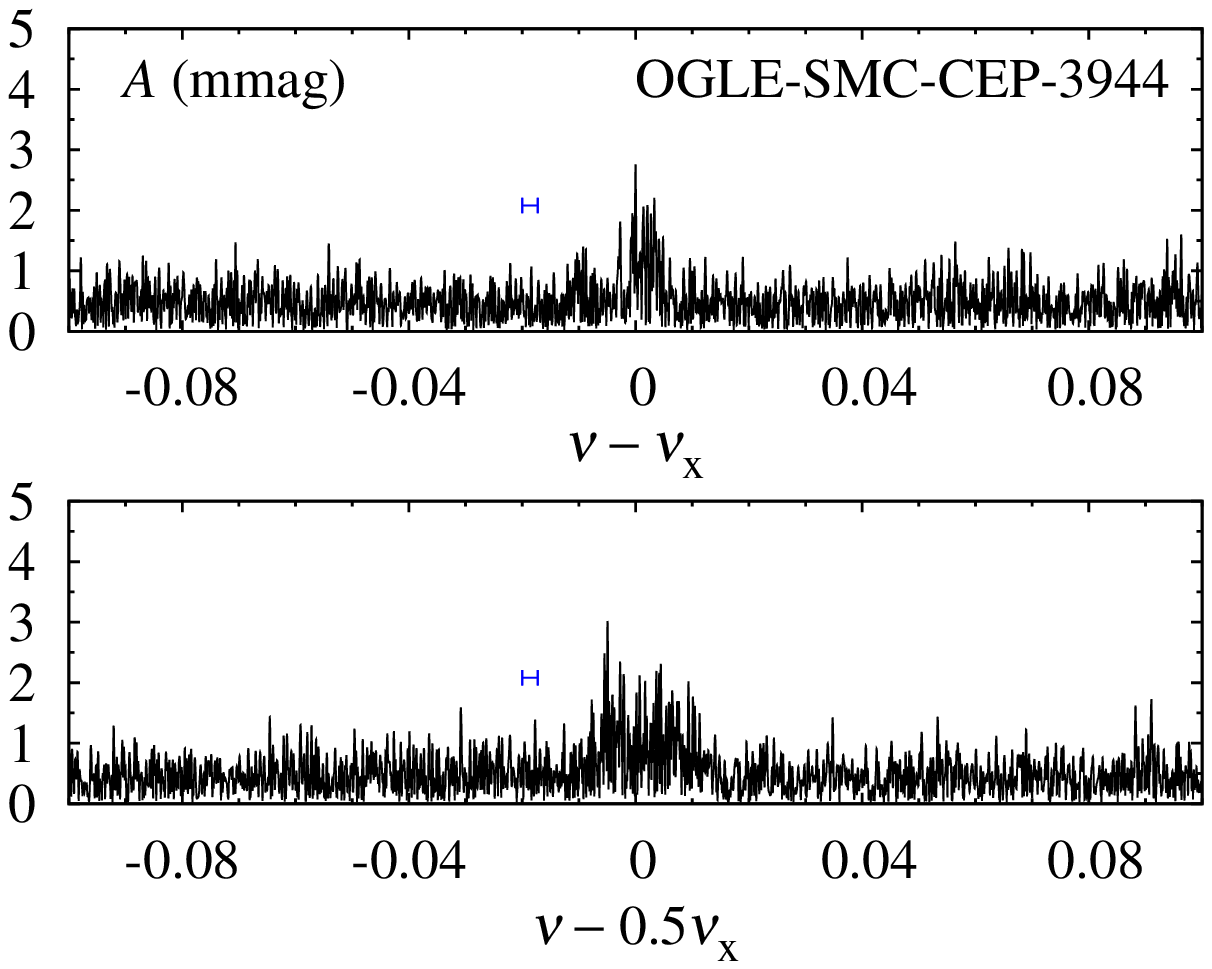}}
\resizebox{0.33\hsize}{!}{\includegraphics{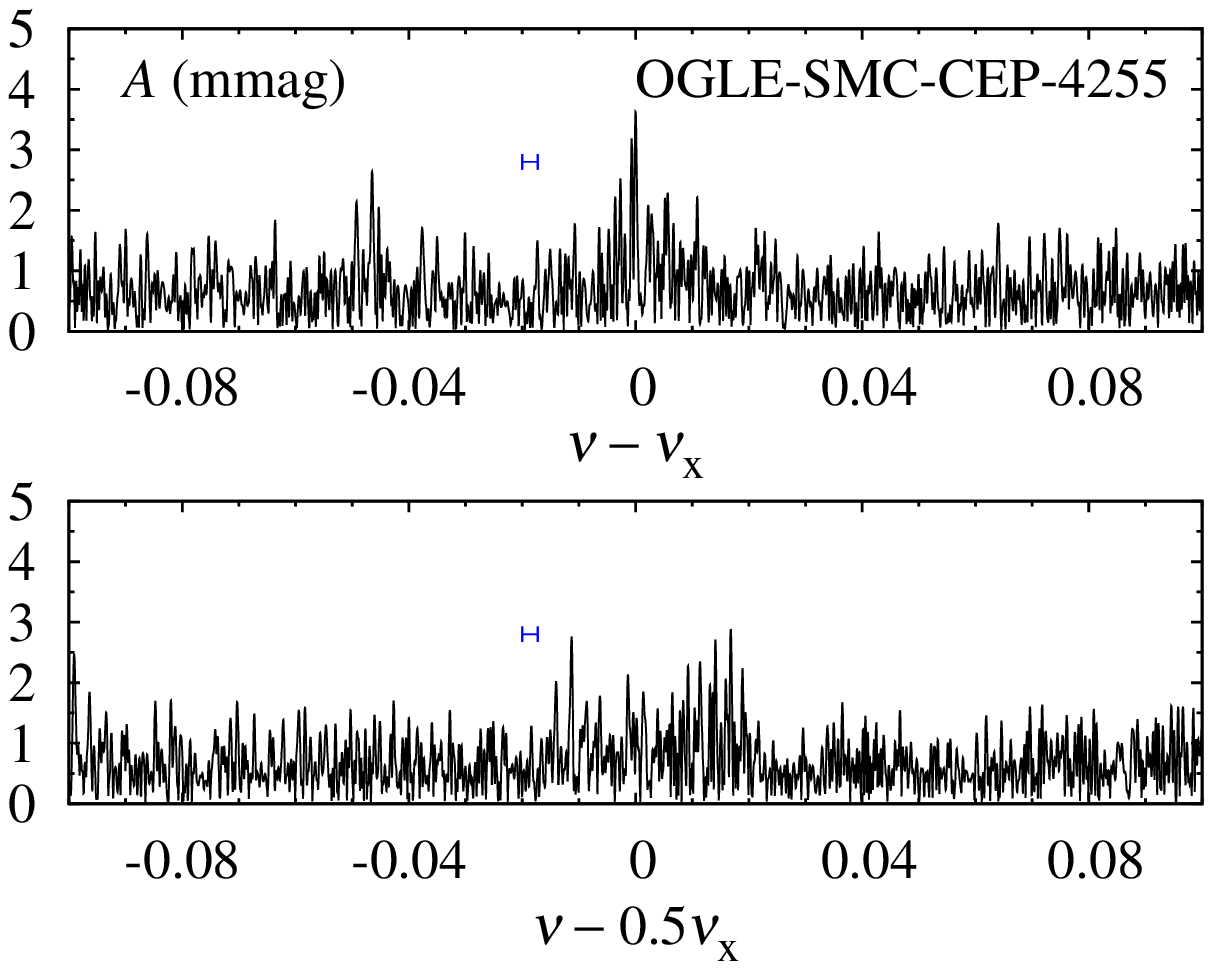}}
\caption{The same as Fig.~\ref{fig:sh_nice}, but for stars with broad power excess at subharmonic frequency.}
\label{fig:sh_broad}
\end{figure*}

\begin{figure*}
\centering
\noindent\resizebox{0.33\hsize}{!}{\includegraphics{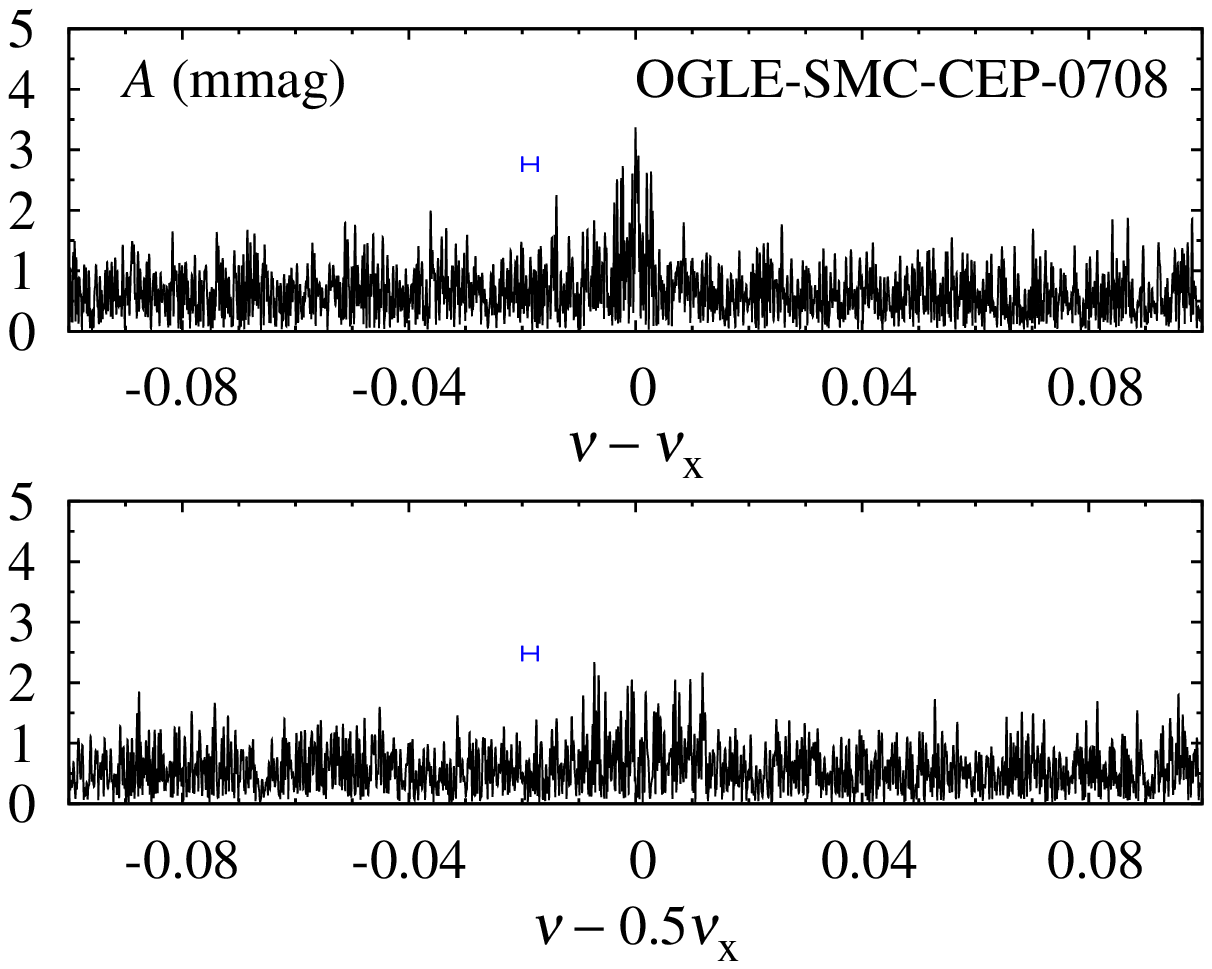}}
\resizebox{0.33\hsize}{!}{\includegraphics{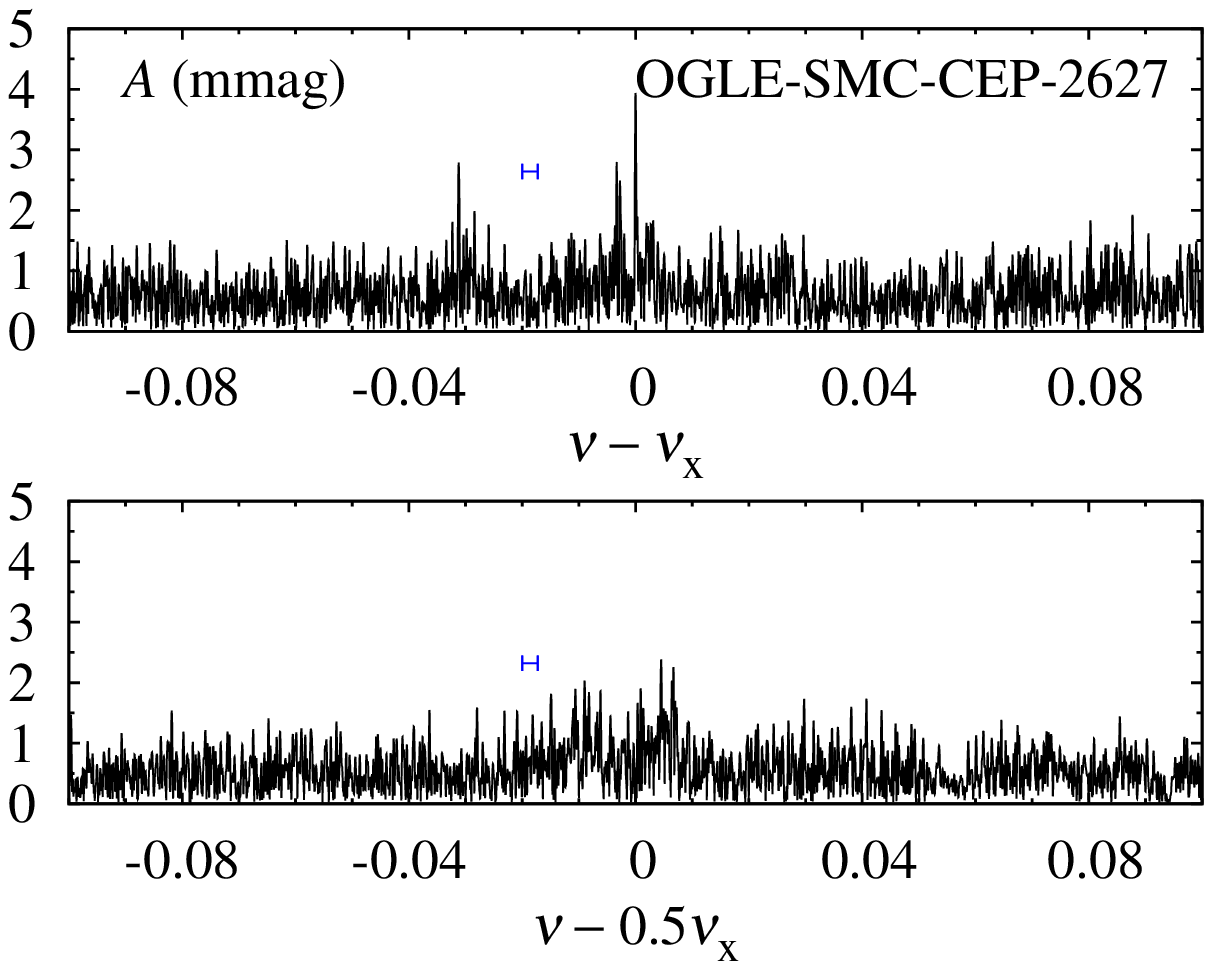}}
\resizebox{0.33\hsize}{!}{\includegraphics{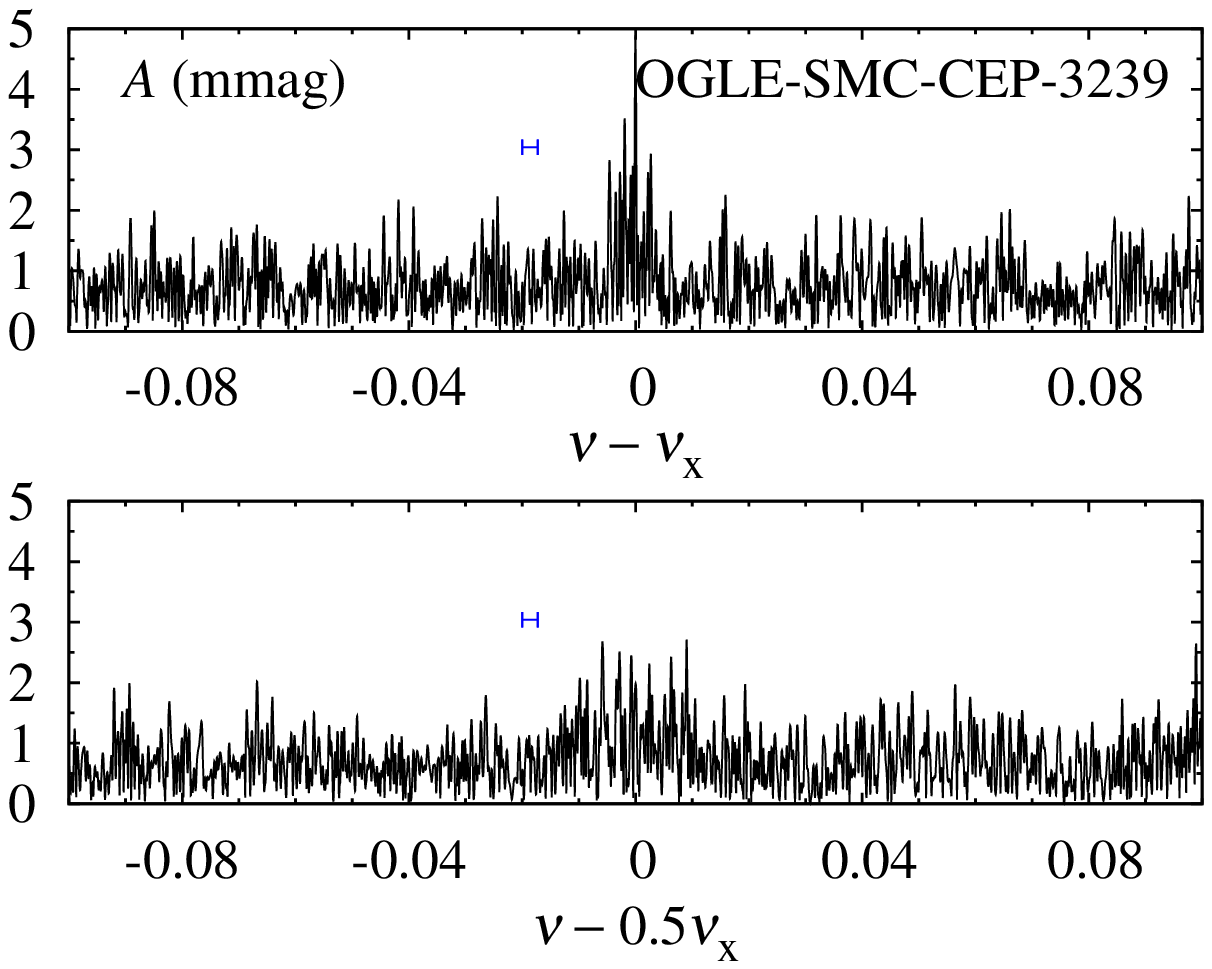}}
\caption{The same as Fig.~\ref{fig:sh_nice}, but for stars with weak detection of power excess at subharmonic frequency.}
\label{fig:sh_weak}
\end{figure*}

\begin{table*}
\caption{Stars with significant power excess centered at sub-harmonic frequency, $1/2\fx$. Consecutive columns contain: star's id, period ratio, $\pxpo$, frequency of the additional variability, $\fx$, frequency of the highest peak detected around $1/2\fx$, $\fsh$, amplitude of the additional variability, $A_{\rm x}$, and amplitude ratio, $A_{\rm sh}/A_{\rm x}$, approximate $S/N$ for the peak at $\fsh$ and remarks: `weak' -- weak detection, `broad' -- broad power excess; `tdp' -- time-dependent prewhitening of all signals except $\fsh$ conducted.}
\label{tab:sh}
\begin{tabular}{lrrrrrrrl}
star & $\pxpo$ & $\fx$\thinspace (1/d) & $\fsh$\thinspace (1/d) & $\fsh/\fx$ & $A_{\rm x}$\thinspace (mag) & $A_{\rm sh}/A_{\rm x}$ & $S/N$ & remarks \\
\hline
OGLE-SMC-CEP-0212  & 0.6247 & 0.91940(4) & 0.46017(4) & 0.5005 & 0.0035 & 0.95 & 4.0 & \\
OGLE-SMC-CEP-0251  & 0.6249 & 0.89065(1) & 0.44722(1) & 0.5021 & 0.0039 & 1.02 & 6.5 & \\ 
OGLE-SMC-CEP-0628  & 0.6231 & 0.87158(2) & 0.42911(2) & 0.4923 & 0.0032 & 1.13 & 6.1 & \\
OGLE-SMC-CEP-0708  & 0.6291 & 1.23406(2) & 0.60976(3) & 0.4941 & 0.0034 & 0.68 & 3.8 & tdp, weak \\ 
OGLE-SMC-CEP-0866  & 0.6223 & 0.92166(2) & 0.45884(2) & 0.4978 & 0.0027 & 0.81 & 4.1 & tdp, weak \\
OGLE-SMC-CEP-1119  & 0.6260 & 0.96230(2) & 0.48410(2) & 0.5031 & 0.0042 & 1.11 & 6.2 & \\  
OGLE-SMC-CEP-1127  & 0.6270 & 0.97348(3) & 0.48376(3) & 0.4969 & 0.0033 & 0.90 & 4.2 & \\
OGLE-SMC-CEP-1366  & 0.6224 & 0.75493(3) & 0.38902(4) & 0.5153 & 0.0034 & 0.68 & 4.1 & tdp, broad \\ 
OGLE-SMC-CEP-1516  & 0.6248 & 0.81994(4) & 0.41672(4) & 0.5082 & 0.0021 & 0.93 & 3.7 & weak \\
OGLE-SMC-CEP-1710  & 0.6217 & 0.76026(1) & 0.37468(2) & 0.4928 & 0.0031 & 0.56 & 4.7 & tdp \\ 
OGLE-SMC-CEP-1771  & 0.6391 & 0.57512(2) & 0.28687(3) & 0.4988 & 0.0022 & 0.59 & 4.1 & tdp, weak \\ 
OGLE-SMC-CEP-1773  & 0.6249 & 0.78102(2) & 0.38892(2) & 0.4980 & 0.0026 & 0.82 & 4.2 & tdp, broad \\ 
OGLE-SMC-CEP-1856  & 0.6223 & 0.84651(1) & 0.43373(2) & 0.5124 & 0.0046 & 0.45 & 4.1 & tdp, broad \\ 
OGLE-SMC-CEP-1870  & 0.6333 & 0.40092(1) & 0.19750(2) & 0.4926 & 0.0026 & 0.52 & 4.5 & tdp \\ 
OGLE-SMC-CEP-1975  & 0.6235 & 0.84083(3) & 0.42308(3) & 0.5032 & 0.0028 & 0.89 & 4.4 & \\
OGLE-SMC-CEP-2178  & 0.6199 & 0.65149(2) & 0.32576(3) & 0.5000 & 0.0029 & 0.84 & 4.9 & \\
OGLE-SMC-CEP-2227  & 0.6207 & 0.77645(2) & 0.39090(2) & 0.5034 & 0.0024 & 0.73 & 4.3 & \\
OGLE-SMC-CEP-2253  & 0.6356 & 0.45460(2) & 0.22936(2) & 0.5045 & 0.0023 & 0.77 & 3.9 & weak \\ 
OGLE-SMC-CEP-2285  & 0.6240 & 0.79324(2) & 0.40705(2) & 0.5131 & 0.0033 & 0.69 & 4.3 & tdp, broad \\ 
OGLE-SMC-CEP-2433  & 0.6409 & 0.65298(2) & 0.32604(2) & 0.4993 & 0.0032 & 0.74 & 4.0 & weak \\ 
OGLE-SMC-CEP-2567  & 0.6207 & 0.71286(3) & 0.35488(3) & 0.4978 & 0.0041 & 0.76 & 4.4 & \\ 
OGLE-SMC-CEP-2594  & 0.6222 & 0.77203(2) & 0.38637(3) & 0.5005 & 0.0043 & 0.90 & 5.7 & \\ 
OGLE-SMC-CEP-2627  & 0.6212 & 0.77597(2) & 0.39251(2) & 0.5058 & 0.0038 & 0.62 & 4.1 & tdp, broad, weak \\ 
OGLE-SMC-CEP-2628  & 0.6228 & 0.91935(2) & 0.45904(3) & 0.4993 & 0.0028 & 0.78 & 3.9 & weak \\
OGLE-SMC-CEP-2681  & 0.6228 & 0.77327(1) & 0.37618(2) & 0.4865 & 0.0047 & 0.57 & 4.1 & broad \\ 
OGLE-SMC-CEP-2805  & 0.6179 & 0.70517(2) & 0.34070(3) & 0.4831 & 0.0036 & 0.83 & 5.1 & \\
OGLE-SMC-CEP-2813  & 0.6204 & 0.71032(1) & 0.35781(2) & 0.5037 & 0.0030 & 0.98 & 5.4 & \\ 
OGLE-SMC-CEP-3040  & 0.6393 & 0.62720(2) & 0.31610(2) & 0.5040 & 0.0025 & 0.67 & 4.0 & weak \\
OGLE-SMC-CEP-3172  & 0.6242 & 0.96361(2) & 0.47473(2) & 0.4927 & 0.0031 & 0.74 & 4.2 & broad \\ 
OGLE-SMC-CEP-3239  & 0.6219 & 0.89478(3) & 0.45643(4) & 0.5101 & 0.0046 & 0.55 & 3.6 & weak \\ 
OGLE-SMC-CEP-3292  & 0.6189 & 0.64590(2) & 0.32966(3) & 0.5104 & 0.0026 & 0.57 & 3.9 & tdp, weak, broad\\ 
OGLE-SMC-CEP-3298  & 0.6389 & 0.58397(2) & 0.29091(2) & 0.4982 & 0.0023 & 0.88 & 5.3 & \\ 
OGLE-SMC-CEP-3312  & 0.6234 & 0.75752(2) & 0.36821(2) & 0.4861 & 0.0029 & 0.79 & 4.3 & broad \\
OGLE-SMC-CEP-3317  & 0.6244 & 0.80069(3) & 0.39738(1) & 0.4963 & 0.0029 & 1.08 & 5.1 & \\ 
OGLE-SMC-CEP-3343  & 0.6250 & 0.87348(1) & 0.43812(2) & 0.5016 & 0.0047 & 0.76 & 6.3 & \\ 
OGLE-SMC-CEP-3349  & 0.6261 & 0.83096(2) & 0.42282(3) & 0.5088 & 0.0031 & 0.81 & 4.3 & broad \\ 
OGLE-SMC-CEP-3590  & 0.6241 & 0.79812(1) & 0.40593(2) & 0.5086 & 0.0027 & 0.62 & 3.9 & tdp, weak, broad \\
OGLE-SMC-CEP-3944  & 0.6232 & 0.74217(2) & 0.36615(2) & 0.4933 & 0.0028 & 1.09 & 6.0 & broad \\ 
OGLE-SMC-CEP-3987  & 0.6350 & 0.46744(2) & 0.23851(3) & 0.5102 & 0.0039 & 0.57 & 4.3 & tdp, weak \\ 
OGLE-SMC-CEP-4011  & 0.6256 & 0.88095(2) & 0.43534(3) & 0.4942 & 0.0035 & 0.98 & 5.4 & \\ 
OGLE-SMC-CEP-4046  & 0.6217 & 0.71527(2) & 0.35977(3) & 0.5030 & 0.0044 & 0.75 & 5.2 &  \\ 
OGLE-SMC-CEP-4068  & 0.6205 & 0.80152(2) & 0.40238(2) & 0.5020 & 0.0034 & 1.03 & 6.2 & broad \\ 
OGLE-SMC-CEP-4205  & 0.6236 & 0.82450(2) & 0.41226(3) & 0.5000 & 0.0059 & 0.64 & 5.5 & \\ 
OGLE-SMC-CEP-4255  & 0.6246 & 0.76557(3) & 0.37150(4) & 0.4853 & 0.0039 & 0.76 & 4.3 & broad\\ 
OGLE-SMC-CEP-4262  & 0.6226 & 0.76736(3) & 0.38411(4) & 0.5006 & 0.0041 & 0.85 & 4.2 & \\
OGLE-SMC-CEP-4388  & 0.6226 & 0.78952(3) & 0.39529(3) & 0.5007 & 0.0045 & 1.11 & 5.7 & \\ 
OGLE-SMC-CEP-4395  & 0.6373 & 0.54682(3) & 0.27416(4) & 0.5014 & 0.0033 & 0.75 & 4.1 & weak \\ 
OGLE-SMC-CEP-4462  & 0.6201 & 0.75481(3) & 0.37636(3) & 0.4986 & 0.0030 & 1.09 & 5.2 & \\ 
\hline
\end{tabular}
\end{table*}

No firm detection of power excess at other subharmonic frequencies, i.e. at $3/2\fx$, $5/2\fx$, etc., is reported. There are a few ambiguous cases, in which power excess is present at $3/2\fx$, but sometimes it is an alias of power excess at $1/2\fx$ or of unresolved residual power at first overtone frequency. Time-dependent prewhitening will not help here; by removing e.g. the non-stationary variation at $\fo$ we also remove the power at its daily aliases.

Before we discuss the properties of the signals detected at subharmonic frequencies, in Figs.~\ref{fig:sh_nice}, \ref{fig:sh_broad} and \ref{fig:sh_weak} we show some examples of structures detected in the frequency spectra of the stars at $\fx$ and at $1/2\fx$. In Fig.~\ref{fig:sh_nice}, we show the cases in which signal at $1/2\fx$ is firmly detected and is relatively narrow. In Fig.~\ref{fig:sh_broad}, we show some of the cases in which the signal at $1/2\fx$ is broad. Finally, in Fig.~\ref{fig:sh_weak}, we show some cases in which detection of the power excess is weak. Structure of these three figures is the same. For each star two panels are plotted. In the top panel the frequency spectrum centred at $\fx$ is plotted, while in the bottom panel the frequency spectrum centred at $1/2\fx$ is plotted. The plotted frequency range is the same in the two panels; it is wider in Figs.~\ref{fig:sh_broad} and \ref{fig:sh_weak} for better visualization of the signal at $1/2\fx$. 

Based on the content of Tab.~\ref{tab:sh} and on Figs.~\ref{fig:sh_nice}, \ref{fig:sh_broad} and \ref{fig:sh_weak}, we now discuss the properties of the signal detected at $1/2\fx$. We first note that the detected power excess is indeed well centred at $1/2\fx$. The mean value of $\fsh/\fx$ for all the stars is $0.5003\pm0.0010$; values of $\fsh/\fx<0.5$ are as common as $\fsh/\fx>0.5$. The values of $|\fsh/\fx-0.5|$ (see Tab.~\ref{tab:sh}) never exceed $0.02$. The largest deviations are present for stars in which broad power excess at subharmonic is observed, like in those plotted in Fig.~\ref{fig:sh_broad} (OGLE-SMC-CEP-1856, -4255). Still, there is no doubt that the power excess is centred at $1/2\fx$ (only the highest peak within the power excess is located a bit off).

Stars that do show power excess at $1/2\fx$ are not uniformly distributed in the Petersen diagram. In Fig.~\ref{fig:pet}, stars which show firm power excess at $1/2\fx$ are plotted with filled symbols, while for stars in which the detection of power excess is weak are marked with half-filled symbols. Majority of the 48 stars with power excess at $1/2\fx$ fall within the middle sequence (40 stars, including 8 weak cases), a few stars fall within the top sequence (8 stars including 6 weak cases) and none falls within the bottom sequence. Thus $74$\thinspace per cent of stars from the middle sequence and $31$\thinspace per cent of the stars from the top sequence show the power excess at $1/2\fx$. If we exclude the weak detections, the numbers are $59$ and $8$\thinspace per cent, respectively. We conclude that the occurrence of power excess at $1/2\fx$ is strongly correlated with the location of star on the Petersen diagram. The power excess is detected in the majority of stars from the middle sequence, in significantly smaller fraction of stars from the top sequence and in no star from the bottom sequence.  

\begin{figure}
\centering
\noindent\resizebox{\hsize}{!}{\includegraphics{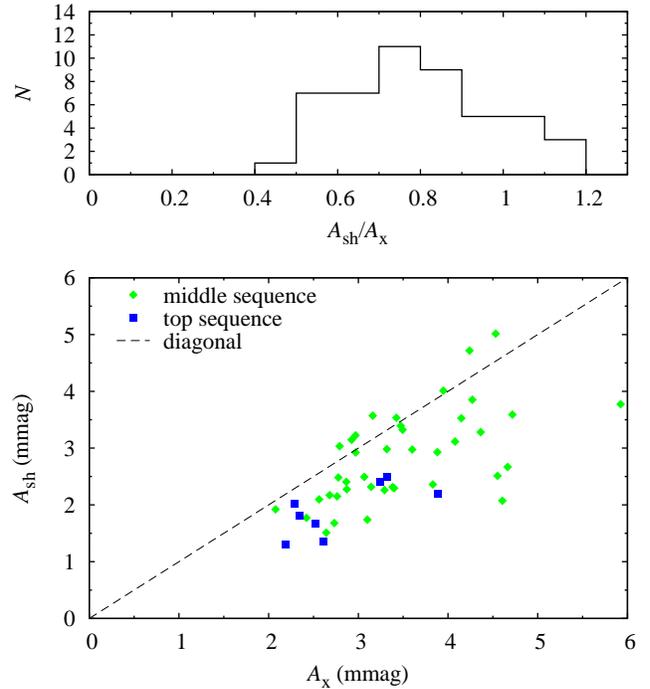}}
\caption{Histogram of amplitude ratios $\ash/\ax$ (top panel) and plot of $\ash$ versus $\ax$ (bottom panel).}
\label{fig:sh_amps}
\end{figure}

The amplitude of the signal at $\fsh$ is always in the mmag range and is comparable to the amplitude of the signal at $\fx$. This is investigated in more detail in Fig.~\ref{fig:sh_amps}, in which in the top panel we plot the histogram of amplitude ratio, $\ash/\ax$, and in the bottom panel we plot $\ash$ versus $\ax$. The distribution of amplitude ratios is wide, without a pronounced peak. It is truncated at $\ash/\ax\approx 0.5$, which is not surprising. As the signals at $\fx$ are weak, detected with typical $S/N\approx 4-6$, we cannot detect the signals with amplitudes significantly lower, below $\approx0.5\ax$ -- these are hidden in the noise. We note eight cases in which peak at $\fsh$ is higher than the peak detected at $\fx$. We may safely conclude that amplitudes, $\ax$ and $\ash$, are comparable. This is further illustrated in the bottom panel of Fig.~\ref{fig:sh_amps}, in which $\ash$ is plotted versus $\ax$. Green diamonds correspond to stars located within the middle sequence in the Petersen diagram, while blue squares correspond to stars located within the top sequence. The amplitudes are (weakly) correlated; the higher the $\ax$, the higher the $\ash$. Amplitudes in stars from the top sequence are in general smaller (see also Sect.~\ref{ssec:selection}).

\subsection{Possible impact of observational selection effects on incidence rates}\label{ssec:selection}

The period distribution of stars with additional variability, and the incidence rate of power excess at subharmonic frequency within each sequence, may be affected by observational selection effects. The most important selection effect is related to star's luminosity. On the short-period end, the Cepheids are least luminous (Fig.~\ref{fig:cwa_basics}). Therefore, we expect larger noise level in the Fourier transform. As a consequence, low-amplitude variability may be hidden in the noise, which could explain the lack of additional variability at $\fx$ in the shortest period first overtone stars or lack of subharmonics in stars of the bottom sequence in the Petersen diagram. On long-period end, the stars are more luminous and the noise level should be lower.

\begin{figure}
\centering
\noindent\resizebox{\hsize}{!}{\includegraphics{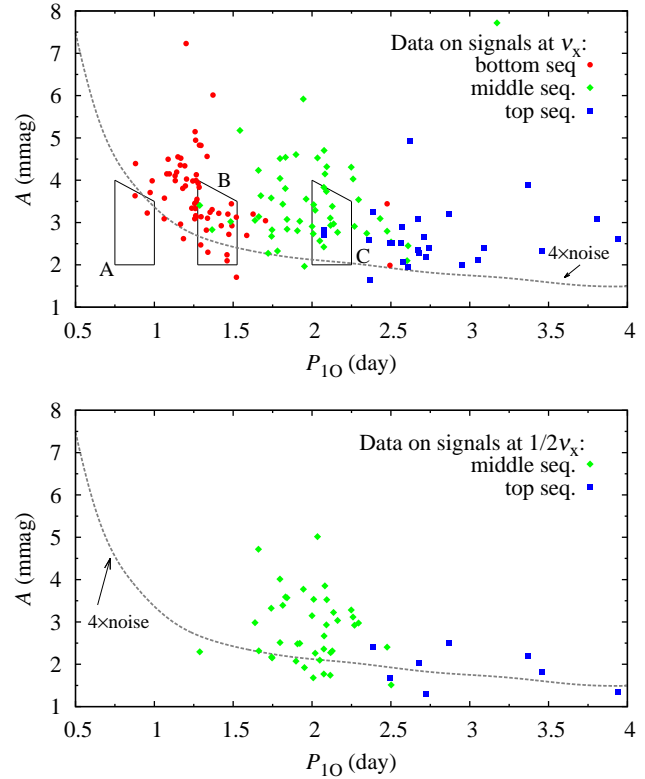}}
\caption{The influence of the observational selection effects on the period distribution of stars with the additional variability at $\fx$ (top panel) and on the incidence rate of power excess at $1/2\fx$ for stars of the three sequences in the Petersen diagram (bottom panel).}
\label{fig:sel}
\end{figure}

To quantify the impact of noise on the detection of low-amplitude variability, we first estimate the noise level in the Fourier transform as a function of pulsation period. To this aim, we analysed the data for all SMC OGLE-III 1O Cepheids in the following, homogeneous way. Using the time-dependent prewhitening on a season-to-season basis we removed from the data (possibly non-stationary) variability associated with the first overtone (sixth order Fourier series). This technique also removes trends possibly present in the data. Possible signals at $\fx$ and at $1/2\fx$ remain in the data but, as these signals are of low amplitude and present in less than 10\thinspace per cent of all 1O Cepheids, they should not alter our estimate significantly. Then, in the frequency spectrum of residual data for each star, we computed the average noise level in the frequency range $(0,\,3\fo)$. The resulting data, noise versus first overtone period, were fitted with spline function. This function, multiplied by 4, is plotted with dashed line in Fig.~\ref{fig:sel} and represents the estimate of the detection threshold as a function of the first overtone period. 
  
In the top panel of Fig.~\ref{fig:sel}, we consider the influence of selection effects on the detection of additional variability at $\fx$. Different symbols represent the data for the three sequences. As noted in Sect.~\ref{ssec:amps} (see also Fig.~\ref{fig:histoX}) the discussed form of pulsation is not present for $\Po<0.75{\rm d}$ and is very scarce in the $ 0.75{\rm d}\!<\!\Po\!<\!1{\rm d}$ range in which 1O Cepheids are very numerous. The question we can answer is, whether the sharp decrease of the incidence rate at shorter periods can be explained by observational selection, assuming that the amplitude distribution of signals at $\fx$ is similar as is observed for longer periods.

For the shortest periods, $\Po\!<\!0.75{\rm d}$, it is clear from Fig.~\ref{fig:sel} that the noise level is very high. It strongly depends on pulsation period, but already at $\Po=0.75{\rm d}$ detection of signals with amplitudes below 4.5\thinspace mmag might be difficult and situation worsens fast as period is decreased further. Thus, the lack of additional variability for the  $\Po\!<\!0.75{\rm d}$ range may be entirely due to selection effects. 

In the $0.75{\rm d}\!<\!\Po\!<\!1{\rm d}$ range the situation is more difficult. We note that for longer periods we can detect signals with amplitudes above $\sim\!2$ mmag. Thus, the very low number of stars with additional variability located within box A in Fig.~\ref{fig:sel}, which is due to high noise level, may be responsible for the small incidence rate in the discussed period range as compared to longer periods. How many detections may we miss in this box? As an estimate, we can count the stars within the same area, but at longer periods, for example within boxes marked B or C in Fig.~\ref{fig:sel}.  There are 21 stars within box B and 15 stars within box C. These numbers are scaled by $N_{\rm A}/N_{\rm B}$ or $N_{\rm A}/N_{\rm C}$ factor, where $N_{\rm X}$ is the total number of 1O Cepheids within period range corresponding to a given box. Using these numbers to estimate the incidence rate within the $0.75{\rm d}\!<\!\Po\!<\!1{\rm d}$ range we get $11.1\pm 1.7$\thinspace per cent or $14.6\pm 1.9$\thinspace per cent, using data from box B or C, respectively. We conclude that sharp decrease of the incidence rate at $\Po<1{\rm d}$ may be entirely explained by observational selection. 

For the period range characteristic for the middle and top sequences in the Petersen diagram, the noise level is roughly the same, it slowly decreases with increasing period. Thus, a small decrease of the incidence rate of the discussed form of pulsation, with pulsation period, noted in Sect.~\ref{ssec:amps}, is likely real and not a result of observational selection.

In the bottom panel of Fig.~\ref{fig:sel}, we consider the influence of observational selection on the incidence rate of power excess at $1/2\fx$ within the three sequences. Symbols correspond to observational data. We first note that amplitudes of signals in stars of the top sequence are, on average, lower than amplitudes of signals in stars of the middle sequence. Also, the incidence rate of power excess at subharmonic is significantly lower for the top sequence. This is obviously not due to different noise levels; signals of the amplitude characteristic for the middle sequence should be easily detected at longer periods. The possible explanation is that amplitudes of the signals at $1/2\fx$ in stars of the top sequence are lower than in stars of the middle sequence. Situation is similar for the stars of the bottom sequence. Although the noise level increases with the decreasing period, the increase is pronounced only for $\Po\lesssim 1{\rm d}$. Amplitudes as high as in the middle sequence should be detected, but they are not. Thus, the amplitudes of signals at $1/2\fx$ in stars of the bottom sequence must be lower than in stars of the middle sequence. In fact, to remain undetectable, they cannot be higher than in the top sequence. The sequence-dependent amplitude of power excess at $1/2\fx$ is in line with the theory proposed recently by \cite{wd16} to explain the discussed form of pulsation. In this theory, the signals at $1/2\fx$ should be observed for all sequences and correspond to non-radial modes of different $\ell$, for which the observed amplitudes differ due to geometric cancellation. We discuss it in more detail in Sect.~\ref{ssec:nature}.

\subsection{Time-variability}\label{ssec:tv}

\begin{figure}
\centering
\resizebox{\hsize}{!}{\includegraphics{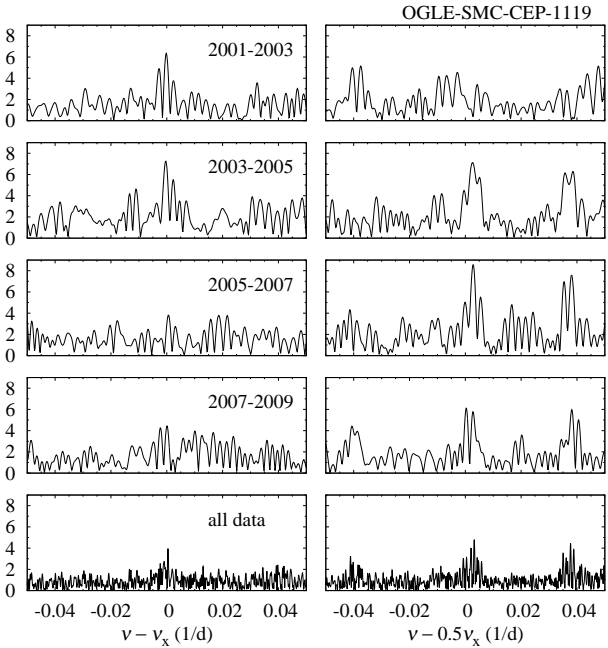}}
\caption{Seasonal analysis of the frequency spectra centred at $\fx$ (left panels) and at $1/2\fx$ (right panels) for OGLE-SMC-CEP-1119. The bottom panels show the frequency spectra for all data. Structure at $\nu-0.5\nu_{\rm x}\approx 0.04$ (in the right panels) is a daily alias of structure centred at $0.5\nu_{\rm x}$.}
\label{fig:s1119}
\end{figure}

\begin{figure}
\centering
\resizebox{\hsize}{!}{\includegraphics{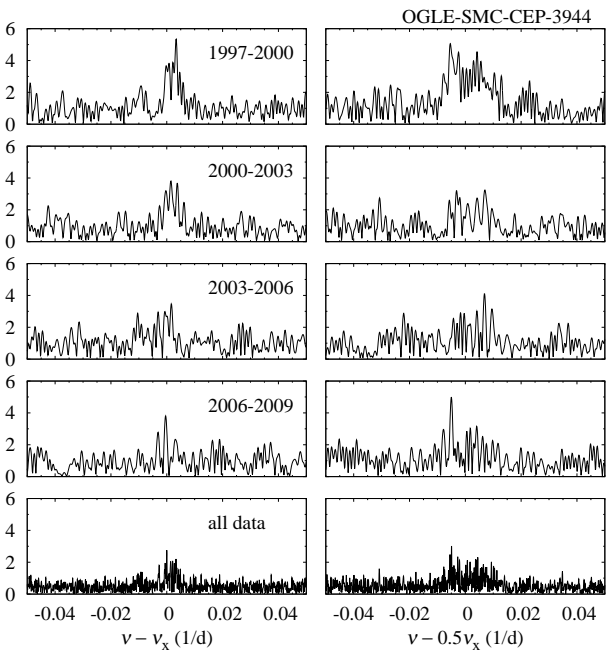}}
\caption{Same as Fig.~\ref{fig:s1119}, but for OGLE-SMC-CEP-3944.}
\label{fig:s3944}
\end{figure}

Complex, non-coherent, and often broad structures, detected both at $\fx$ and in some stars at $1/2\fx$ (Figs.~\ref{fig:sh_nice}, \ref{fig:sh_broad} and \ref{fig:sh_weak}), indicate strong time variation of the amplitude and/or phase of the variability these structures represent. Because of the low amplitudes of these signals, typically between 2 and 5 mmags, and relatively high noise level in the ground-based observations, it is difficult to analyse this variability in more detail, with high time resolution. Still, some analysis is possible for stars in which the signals are detected at relatively high $S/N$. For these stars we divided the data into four groups, each consisting of two (or in some cases three) observing seasons. Then for each group we calculated the discrete Fourier transform and investigated the frequency spectrum at around $\fx$ and $1/2\fx$. Results of the analysis are presented in Fig.~\ref{fig:s1119} (for OGLE-SMC-CEP-1119) and in Fig.~\ref{fig:s3944} (for OGLE-SMC-CEP-3944). Results are qualitatively similar for a few other stars for which such analysis was possible. 

In Fig.~\ref{fig:s1119}, we observe that from season to season the amplitude and location of the peaks present at $\fx$ and at $1/2\fx$ strongly vary. In particular, the signal at $\fx$ is significant in 2001--2005 seasons, while it is not significant later on. The signal at $1/2\fx$ was insignificant in the first observing seasons, while it was clearly present starting from 2003 and later on. 

For OGLE-SMC-CEP-3944, analysed in Fig.~\ref{fig:s3944}, we observe  that signal at $\fx$ is always present, but its amplitude and/or phase clearly vary. The signal at $1/2\fx$ is weakly marked in between 2000 and 2006. Structure of the observed broad power excess vary in time. As a result, broad and essentially flat power excess is present in the analysis of all data.

\section{Discussion}

\subsection{Comparison with RR~Lyr stars}\label{ssec:rrlcomp}

\begin{figure}
\centering
\resizebox{\hsize}{!}{\includegraphics{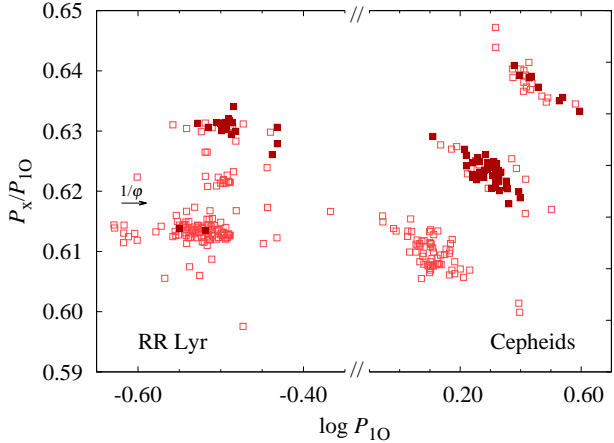}}
\caption{Petersen diagram with RRc stars with additional periodicity from Netzel et al. (2015b) and 1O Cepheids from the present study. Filled symbols correspond to stars with power excess detected at $1/2\fx$. Arrow indicates the reciprocal of the golden ratio.}
\label{fig:rrce}
\end{figure}

A very similar form of variability is detected in RRc stars, as already mentioned in the Introduction. The obvious similarity is the characteristic period ratio, which falls into similar range, $P/\Po\in(0.60,\,0.65)$, and the occurrence of the additional variability in first overtone stars (or in double-mode stars, with fundamental and first overtone modes simultaneously excited). The present study allow a more detailed comparison.
\begin{itemize}

\item {\bf Incidence rate of the phenomenon.} The phenomenon is common among RRc stars; space observations leave no doubt; 14 out of 15 RRc/RRd stars observed from space show the phenomenon \citep[see][]{pamsm15}. Incidence rates in the top-quality ground-based observations are also high, 27\thinspace per cent in the Galactic bulge sample of RRc stars studied by \cite{netzel3} and 38\thinspace per cent in the M3 sample observed by \cite{jurcsik_M3}. Unfortunately, for Cepheids we lack systematic analysis of large sample of stars. The 138 stars considered here constitute $8.4$\thinspace per cent of the OGLE-III sample of 1O SMC Cepheids. The incidence rate depends on the pulsation period; in particular the phenomenon does not occur in the shortest period ($\Po\lesssim 0.8$\thinspace d) Cepheids. For longer periods the incidence rate is $\approx\!8-15$\thinspace per cent. Space observations of 1O Cepheids are scarce. Polaris was observed with star tracker on board {\it Coriolis} satellite \citep{bruntt}, SZ~Tau was observed with {\it MOST} \citep{evans_MOST} and 2 other stars were observed with {\it CoRoT} \citep{poretti}. In these stars additional variability was not detected. Two stars observed with {\it K2} show the discussed form of pulsation (Plachy et al. in prep.).

\item {\bf The Petersen diagram.} Both for Cepheids and for RR~Lyr stars three sequences are present -- see Fig.~\ref{fig:rrce}. The Cepheid sequences are slanted and well separated. The bottom sequence is most populated, but the middle and top sequences are well represented, too. In the case of RR~Lyr stars, the sequences are nearly horizontal, not that well separated and majority of stars fall within the bottom sequence. In both groups stars that belong to more than one sequence are found. While in the case of RR~Lyr stars there are very good examples of stars that belong to three sequences simultaneously \citep[see fig.~5 in][]{netzel3}, in the case of Cepheids only stars that belong to two sequences are found, and these are rather weak cases (Fig.~\ref{fig:2seq}).

\item{\bf Amplitudes.} Both in Cepheids and in RR~Lyr stars the additional periodicity is of low amplitude, in the mmag range. In both cases the most typical amplitude is around 2\thinspace per cent of the first overtone amplitude [compare fig.~7 in \cite{netzel3} and Fig.~\ref{fig:hia}].

\item{\bf Subharmonics.} Both in Cepheids and in RR~Lyr stars significant power excess centred at $1/2\fx$ is detected. Subharmonics are detected in 20\thinspace per cent of the sample of Galactic bulge RRc stars with additional variability analysed by \cite{netzel3} and in 35\thinspace per cent of the present Cepheid sample. \cite{netzel3} reported the power excess also at $3/2\fx$, other cases are known from space observations \citep[see e.g.][]{pamsm15,molnar,kurtz}. No firm detection of power excess at $3/2\fx$ is found in the present Cepheid sample. Stars with power excess at $1/2\fx$, both RR~Lyr and Cepheids, are not uniformly distributed in the Petersen diagram. This is illustrated in Fig.~\ref{fig:rrce} in which we plot the stars from the present sample of SMC Cepheids and from the study of \cite{netzel3} \citep[see also][]{netzel_vise}; stars with power excess at subharmonic are marked with filled symbols.

\item{\bf Time variability of the additional signals.} Both for Cepheids and for RR~Lyr stars the signals at $\fx$ and at $1/2\fx$  are complex: power excesses, or clusters of peaks (sometimes very broad) are detected, rather than single and coherent peaks. Compare fig.~11 in \cite{netzel3} and 2 in \cite{netzel_vise} with Figs.~\ref{fig:sh_nice}, \ref{fig:sh_broad} and \ref{fig:sh_weak}. Such complex structures present in the frequency spectrum correspond to strong and irregular variability of signal's phase and/or amplitude. 

\end{itemize} 

\subsection{Nature of the additional variability}\label{ssec:nature}

Based on the just presented comparison we conclude, that the double-periodic pulsation observed both in RRc stars and in 1O Cepheids, with characteristic period ratios, $\pxpo\in(0.60,\, 0.65)$, is qualitatively very similar. Consequently, nature of the additional variability and mechanism of its excitation are most likely the same. In both cases it cannot be pulsation in two radial modes \citep{wdrs,pamsm15}. A model or theory common for the two groups should be searched for.

\cite{wd12} proposed an explanation for Cepheids. The additional variability at $\fx$ was interpreted as a non-radial f-mode of high angular degree. The three Cepheid sequences were reproduced assuming that additional modes have $\ell=42$ (top sequence) $\ell=46$ (middle sequence) and $\ell=50$ (bottom sequence). Due to geometric cancellation, modes of such large angular degree are expected to have very low observed amplitudes. Two problems arise then, however. First, the geometric cancellation is lower for even-$\ell$ modes and, at high degrees, is roughly the same for these modes (depends only on mode parity). Hence, if $\ell=42$, $46$ and $50$ are observed, then $\ell=44$ and $\ell=48$ should be observed as well (all these modes are linearly unstable), which is not the case. Second, the required intrinsic amplitudes are very high, implying large broadening of the spectral lines. Both problems are discussed in \cite{wd12}. When this model was proposed it was not known that in these stars subharmonics are detected. RR~Lyr stars were not studied by \cite{wd12}.
 
\cite{golden} focused on RR~Lyr stars, in particular on one star observed by {\it Kepler} (KIC5520878) and noticed that its period ratio is close to the reciprocal of the golden ratio, $1/\varphi\approx 0.618$. \cite{golden} argue that the dynamics driven by two frequencies in the golden ratio maximally resist perturbations. As noted by \cite{rs_iau} and also well visible in Fig.~\ref{fig:rrce}, in which we mark $1/\varphi$ with an arrow, the stars avoid the golden ratio. In our opinion the proximity of period ratio to $1/\varphi$ in some RR~Lyr stars is a pure coincidence.

Recently, \cite{wd16} proposed a new explanation, in which the signal observed at $1/2\fx$ corresponds to non-radial, $\ell=7-9$ modes. The signal at $\fx$ is its harmonic, which gains the significant amplitude due to non-linear, quadratic effect. According to this model, stars in which subharmonics are detected should not be distributed uniformly among sequences visible in the Petersen diagram. As geometric cancellation is lower for even-$\ell$ modes we should observe the power excess at subharmonic, i.e. we should detect the true non-radial mode, preferentially in sequences which correspond to even-$\ell$ pulsation. In the case of Cepheids it is the middle sequence ($\ell=8$; $\fx=2\nu_8$). The top sequence corresponds to $\ell=7$, while the lower sequence to $\ell=9$. Geometric cancellation is slightly lower for $\ell=7$ than for $\ell=9$. In Fig.~\ref{fig:rrce}, we observe a nice qualitative confirmation of this theory for Cepheids.

In the case of RR~Lyr stars, the top sequence corresponds to even $\ell$ ($\ell=8$) and subharmonic detection (non-radial $\ell=8$ mode detection) should be more common there. According to the \cite{wd16} theory, the bottom, most populated sequence for RR~Lyr stars, corresponds to odd-$\ell$ mode ($\ell=9$). The weakly marked middle sequence corresponds to the combination frequency ($\nu_8+\nu_9$) and no subharmonics should be detected there. This is indeed what we observe -- Fig.~\ref{fig:rrce}. Detailed description of the model is in preparation (Dziembowski \& Smolec, in prep.)

\section{Summary}

We have analysed 138 1O Cepheids from the SMC in which \cite{ogle_cep_smc} reported additional variability with $P/\Po\in(0.60,\, 0.65)$. These stars form three sequences in the Petersen diagram. Our most important findings are the following.
\begin{itemize}

\item The three sequences in the Petersen diagram are not equally populated. In 64 stars we detect periodicities corresponding to the bottom sequence, in 54 stars corresponding to the middle sequence and in 26 stars corresponding to the top sequence. The numbers do not add up to 138 as in the frequency spectra of six stars we detect two significant periodicities that correspond to two of the three sequences. 

\item The additional variability is always of low amplitude, typically about 2--4\thinspace per cent of the first overtone amplitude (between 2 and 5\thinspace mmag) -- Figs.~\ref{fig:amps} and \ref{fig:hia}.

\item In 35\thinspace per cent of stars (25\thinspace per cent if weak cases are excluded), we detect power excess centred at $1/2\fx$, i.e. at subharmonic. The power excess is often broad and of complex structure (Figs.~\ref{fig:sh_nice} and \ref{fig:sh_broad}). Amplitude of the signal detected at $1/2\fx$ is comparable to amplitude of the signal detected at $\fx$ -- Fig.~\ref{fig:sh_amps}. 

\item Stars in which power excess at subharmonic is detected are not uniformly distributed in the Petersen diagram -- Figs.~\ref{fig:pet} and \ref{fig:rrce}. Subharmonics are detected most frequently in the stars of the middle sequence (in 74\thinspace per cent of stars; 59\thinspace per cent excluding weak cases). Incidence rate is much lower for the top sequence (31\thinspace per cent or 8\thinspace per cent without weak cases). Subharmonics are not detected in stars of the bottom sequence. 

\item The additional variability (both at $\fx$ and at $1/2\fx$) is strongly non-stationary; its amplitude and/or phase, strongly vary in time (Figs.~\ref{fig:s1119} and \ref{fig:s3944}).

\item A similar form of pulsation, in which radial first overtone and additional low-amplitude variability is detected with period ratios $\pxpo\in(0.60,\,0.65)$, is also present in RR~Lyr stars. A detailed comparison we have done indicates, that the nature of this phenomenon is most likely the same in both groups of classical pulsators. Therefore, a common theory to explain this puzzling form of pulsation should be searched for.

\item In the Petersen diagram, the distribution of stars in which power excess at subharmonic is detected is not uniform, both for Cepheids and for RR~Lyr stars (Fig.~\ref{fig:rrce}). The observed distribution favors the model proposed recently by \cite{wd16} (Sect.~\ref{ssec:nature}).

\end{itemize}

It is very important to establish the incidence rate of the discussed form of pulsation in Cepheids, to check whether it is as common as in RR~Lyr stars. The systematic search in the OGLE data was started. It seems crucial however, to observe 1O Cepheids from space with {\it K2} mission. High-precision photometry it gathers, offers the possibility to detect periodicities of very low amplitude. Detection of power excess at subharmonic frequencies for stars that belong to different sequences, and study of the amplitude distribution of these signals, is crucial to test the model proposed by \cite{wd16}.

\section*{Acknowledgements}
This research is supported by the Polish National Science Centre through grants DEC-2012/05/B/ST9/03932 and DEC-2015/17/B/ST9/03421. We are grateful to Pawe\l{} Moskalik for detailed reading of the manuscript and many comments that significantly improved its content. Fruitful and stimulating discussions Wojtek Dziembowski are acknowledged. We acknowledge the summer student program at Nicolaus Copernicus Astronomical Center during which part of this work was completed.


\appendix

\section{Stars with additional variability}\label{app:table}
Table~\ref{tab:atab} list all the stars analysed in the present study and contains their basic properties.

\begin{table*}
\caption{Properties of first overtone Cepheids with additional variability. The consecutive columns contain: star's id, first overtone period, $\Po$, period of the additional variability, $\Px$, period ratio, $\pxpo$, amplitude of the first overtone, $A_{\rm 1O}$, and amplitude ratio, $A_{\rm x}/A_{\rm 1O}$, and remarks: `al' -- daily alias of signal at $\fx$ is higher; `nsx' -- complex appearance of the signal at $\fx$; `nsO' -- non-stationary first overtone; `cf' -- combination frequency of $\fx$ and $\fo$ detected; `sh' -- power excess at sub-harmonic frequency (around $1/2\fx$) detected (`{\it sh}' -- weak detection); `ap' -- additional periodicity detected; `tdp' -- time-dependent analysis was conducted; `?' -- weak detection ($S/N$ given in the parenthesis).}
\label{tab:atab}
\begin{tabular}{lrrrrrr}
star & $\Po$\thinspace (d) & $\Px$\thinspace (d) & $\pxpo$ & $A_{\rm 1O}$\thinspace (mag) & $A_{\rm x}/A_{\rm 1O}$ & remarks \\
\hline
OGLE-SMC-CEP-0056  & 0.9860208(7) & 0.60373(1) & 0.6123 & 0.1689 & 0.024 & ? ($S/N=3.77$) \\
OGLE-SMC-CEP-0212  & 1.741010(4)  & 1.08766(4) & 0.6247 & 0.0997 & 0.036 & sh, nsx \\ 
OGLE-SMC-CEP-0251  & 1.796802(1)  & 1.12279(2) & 0.6249 & 0.1399 & 0.029 & sh, nsx \\ 
OGLE-SMC-CEP-0280  & 1.675191(1)  & 1.04344(2) & 0.6229 & 0.1377 & 0.026 & nsO, ap \\
OGLE-SMC-CEP-0281  & 1.2662457(7) & 0.774075(9)& 0.6113 & 0.1263 & 0.033 & al, nsx \\ 
OGLE-SMC-CEP-0307  & 0.9734743(7) & 0.59718(1) & 0.6134 & 0.1922 & 0.019 & nsO, ap \\
OGLE-SMC-CEP-0447  & 1.2651448(8) & 0.77624(1) & 0.6136 & 0.1300 & 0.024 & nsO, nsx \\ 
OGLE-SMC-CEP-0456  & 0.8776927(4) & 0.540588(7)& 0.6159 & 0.1579 & 0.023 & nsO, tdp, ? ($S/N=3.93$) \\
OGLE-SMC-CEP-0466  & 1.1317607(5) & 0.690278(8)& 0.6099 & 0.1603 & 0.026 & al \\
OGLE-SMC-CEP-0491  & 1.0641278(4) & 0.652731(8)& 0.6134 & 0.1679 & 0.021 &    \\
OGLE-SMC-CEP-0509  & 1.626176(1)  & 0.98494(2) & 0.6057 & 0.1322 & 0.024 & ap \\
OGLE-SMC-CEP-0592  & 1.1495482(9) & 0.702313(9)& 0.6109 & 0.1371 & 0.033 & al, nsO, nsx \\ 
OGLE-SMC-CEP-0628  & 1.841217(1)  & 1.14733(2) & 0.6231 & 0.1385 & 0.022 & sh, nsx, nsO \\
OGLE-SMC-CEP-0631  & 2.346706(4)  & 1.46774(6) & 0.6254 & 0.1253 & 0.022 & al, ap \\ 
OGLE-SMC-CEP-0680  & 1.483925(1)  & 0.93037(2) & 0.6270 & 0.1296 & 0.023 & nsO \\
OGLE-SMC-CEP-0696  & 2.359034(2)  & 1.51060(4) & 0.6403 & 0.1065 & 0.024 & nsO, nsx\\ 
OGLE-SMC-CEP-0708  & 1.2881769(6) & 0.81034(1) & 0.6291 & 0.1738 & 0.020 & al, {\it sh}, nsx \\ 
OGLE-SMC-CEP-0759  & 1.2615578(6) & 0.770574(7)& 0.6108 & 0.1450 & 0.034 & nsO, nsx \\
OGLE-SMC-CEP-0797  & 2.623163(3)  & 1.67174(2) & 0.6373 & 0.0840 & 0.059 & al, nsO, nsx, cf, ap\\
OGLE-SMC-CEP-0800  & 1.2598153(8) & 0.77040(2) & 0.6115 & 0.1683 & 0.020 &  \\
OGLE-SMC-CEP-0828  & 1.1798295(9) & 0.714330(9)& 0.6055 & 0.1040 & 0.037 & nsO \\
OGLE-SMC-CEP-0833  & 1.1328767(6) & 0.69237(1) & 0.6112 & 0.1467 & 0.027 & nsO \\
OGLE-SMC-CEP-0841  & 1.3535878(8) & 0.82278(1) & 0.6079 & 0.1495 & 0.022 & nsx \\ 
OGLE-SMC-CEP-0844  & 1.2560636(5) & 0.77035(1) & 0.6133 & 0.1488 & 0.023 & nsx \\ 
OGLE-SMC-CEP-0866  & 1.743531(1)  & 1.08500(3) & 0.6223 & 0.1205 & 0.022 & al, nsx, {\it sh} \\
OGLE-SMC-CEP-1049  & 1.2042213(6) & 0.736700(9)& 0.6118 & 0.1459 & 0.028 & nsx, nsO \\ 
OGLE-SMC-CEP-1053  & 1.2779760(9) & 0.77804(1) & 0.6088 & 0.1460 & 0.027 &  \\
OGLE-SMC-CEP-1059  & 1.2700610(7) & 0.77623(1) & 0.6112 & 0.1570 & 0.023 & nsO  \\
OGLE-SMC-CEP-1119  & 1.659938(2)  & 1.03918(2) & 0.6260 & 0.1327 & 0.032 & sh, nsO, nsx, cf \\ 
OGLE-SMC-CEP-1127  & 1.638464(2)  & 1.02725(4) & 0.6270 & 0.1315 & 0.023 & sh, nsx, ap \\ 
OGLE-SMC-CEP-1138  & 1.4920903(9) & 0.91313(2) & 0.6120 & 0.1420 & 0.021 & nsO, nsx \\
OGLE-SMC-CEP-1185  & 1.1927515(9) & 0.72660(1) & 0.6092 & 0.1644 & 0.026 &  \\
OGLE-SMC-CEP-1248  & 1.1662128(5) & 0.70943(1) & 0.6083 & 0.1559 & 0.019 &  \\
OGLE-SMC-CEP-1294  & 1.3369590(8) & 0.82015(1) & 0.6134 & 0.1443 & 0.022 & ap \\
OGLE-SMC-CEP-1317  & 1.284214(1)  & 0.78028(1) & 0.6076 & 0.1042 & 0.037 & nsO \\
OGLE-SMC-CEP-1366  & 2.128286(4)  & 1.32463(5) & 0.6224 & 0.0912 & 0.037 & sh, nsO  \\
OGLE-SMC-CEP-1482  & 1.3395479(8) & 0.81439(1) & 0.6080 & 0.1004 & 0.023 & nsO, ap  \\
OGLE-SMC-CEP-1496  & 1.334207(1)  & 0.81109(1) & 0.6079 & 0.1444 & 0.032 & cf, nsx \\ 
OGLE-SMC-CEP-1516  & 1.951989(5)  & 1.21960(6) & 0.6248 & 0.0543 & 0.036 & {\it sh}, ap, nsx \\ 
OGLE-SMC-CEP-1583  & 3.80844(1)   & 2.4164(1)  & 0.6345 & 0.0929 & 0.033 & nsx \\ 
OGLE-SMC-CEP-1710  & 2.115741(1)  & 1.31534(2) & 0.6217 & 0.1237 & 0.025 & sh, nsx, nsO \\ 
OGLE-SMC-CEP-1771  & 2.720656(3)  & 1.73877(5) & 0.6391 & 0.0963 & 0.023 & nsx, {\it sh} \\
OGLE-SMC-CEP-1773  & 2.048935(2)  & 1.28038(3) & 0.6249 & 0.1102 & 0.023 & sh, nsO, nsx \\   
OGLE-SMC-CEP-1783  & 1.087583(1)  & 0.66932(1) & 0.6154 & 0.1418 & 0.032 & ap, nsx \\  
OGLE-SMC-CEP-1802  & 2.608860(3)  & 1.62283(5) & 0.6220 & 0.1043 & 0.024 & nsx \\ 
                   & 2.608860(3)  & 1.66777(7) & 0.6393 & 0.1043 & 0.019 &  \\
OGLE-SMC-CEP-1856  & 1.898340(1)  & 1.18132(1) & 0.6223 & 0.1320 & 0.035 & cf, sh, nsx \\ 
OGLE-SMC-CEP-1870  & 3.938740(6)  & 2.49426(8) & 0.6333 & 0.0980 & 0.027 & nsO, cf, sh, nsx \\ 
OGLE-SMC-CEP-1882  & 1.296622(1)  & 0.79454(1) & 0.6128 & 0.1089 & 0.029 & tdp, nsO, nsx \\ 
OGLE-SMC-CEP-1975  & 1.907327(2)  & 1.18931(4) & 0.6235 & 0.1477 & 0.019 & sh \\
OGLE-SMC-CEP-1976  & 1.2643772(8) & 0.76772(1) & 0.6072 & 0.1421 & 0.028 & nsO, nsx, tdp \\ 
OGLE-SMC-CEP-2009  & 1.962690(2)  & 1.21787(2) & 0.6205 & 0.1118 & 0.032 & nsx \\ 
OGLE-SMC-CEP-2033  & 1.4087293(8) & 0.85869(1) & 0.6095 & 0.1430 & 0.023 & nsx, nsO \\ 
OGLE-SMC-CEP-2047  & 1.2436705(7) & 0.75572(1) & 0.6076 & 0.1480 & 0.027 & nsx \\ 
OGLE-SMC-CEP-2116  & 1.488366(2)  & 0.90331(3) & 0.6069 & 0.1394 & 0.025 & nsx, nsO \\ 
OGLE-SMC-CEP-2126  & 1.371930(1)  & 0.83387(2) & 0.6078 & 0.1530 & 0.039 & al  \\
OGLE-SMC-CEP-2131  & 2.602695(3)  & 1.60403(5) & 0.6163 & 0.1363 & 0.015 & al, nsO \\ 
OGLE-SMC-CEP-2178  & 2.476222(4)  & 1.48924(4) & 0.6014 & 0.1114 & 0.031 & nsx \\
                   & 2.476222(4)  & 1.53494(6) & 0.6199 & 0.1114 & 0.025 & sh \\
\end{tabular}
\end{table*}

\begin{table*}
\contcaption{}
\begin{tabular}{lrrrrrr}
star & $\Po$ (d) & $\Px$ (d) & $\pxpo$ & $A_{\rm 1O}$ & $A_{\rm x}/A_{\rm 1O}$ & remarks \\
\hline
OGLE-SMC-CEP-2227  & 2.075033(1)  & 1.28790(3) & 0.6207 & 0.1289 & 0.019 & al, sh, nsx \\ 
OGLE-SMC-CEP-2253  & 3.460665(5)  & 2.19975(9) & 0.6356 & 0.1011 & 0.023 & {\it sh}, nsx, cf \\
OGLE-SMC-CEP-2285  & 2.020254(3)  & 1.26065(3) & 0.6240 & 0.0864 & 0.038 & nsO, tdp, sh, nsx \\ 
OGLE-SMC-CEP-2299  & 1.201553(2)  & 0.73048(2) & 0.6079 & 0.1535 & 0.047 & \\ 
OGLE-SMC-CEP-2374  & 1.0789226(7) & 0.65798(1) & 0.6099 & 0.1751 & 0.024 & \\
OGLE-SMC-CEP-2433  & 2.389605(2)  & 1.53145(4) & 0.6409 & 0.1160 & 0.028 & {\it sh}, nsx \\
OGLE-SMC-CEP-2440  & 1.830655(2)  & 1.14310(3) & 0.6244 & 0.1452 & 0.026 & nsx, ap \\
OGLE-SMC-CEP-2497  & 1.1847874(5) & 0.72476(1) & 0.6117 & 0.1488 & 0.018 & al, nsO, nsx \\ 
OGLE-SMC-CEP-2499  & 1.781123(1)  & 1.11333(2) & 0.6251 & 0.1369 & 0.017 & cf, nsO, nsx \\
OGLE-SMC-CEP-2501  & 1.2583171(9) & 0.76452(1) & 0.6076 & 0.1629 & 0.032 & cf \\
OGLE-SMC-CEP-2528  & 1.4609104(8) & 0.88543(2) & 0.6061 & 0.1417 & 0.016 & tdp, nsO\\
OGLE-SMC-CEP-2536  & 2.368486(2)  & 1.51326(6) & 0.6389 & 0.1333 & 0.012 & nsO, tdp, ? ($S/N=3.90$) \\
OGLE-SMC-CEP-2567  & 2.260065(5)  & 1.40279(5) & 0.6207 & 0.0983 & 0.041 & sh, nsx \\ 
OGLE-SMC-CEP-2593  & 1.2371639(9) & 0.75189(2) & 0.6078 & 0.1568 & 0.021 & \\
OGLE-SMC-CEP-2594  & 2.081922(4)  & 1.29530(3) & 0.6222 & 0.1238 & 0.034 & sh, cf, nsx \\ 
OGLE-SMC-CEP-2595  & 1.2649697(6) & 0.76775(1) & 0.6069 & 0.1475 & 0.024 & nsx \\ 
OGLE-SMC-CEP-2597  & 2.572680(3)  & 1.64169(5) & 0.6381 & 0.1149 & 0.025 & nsx \\ 
OGLE-SMC-CEP-2627  & 2.074554(2)  & 1.28870(3) & 0.6212 & 0.1194 & 0.032 & al, nsx, {\it sh}\\ 
                   & 2.074554(2)  & 1.34270(4) & 0.6472 & 0.1194 & 0.023 & \\
OGLE-SMC-CEP-2628  & 1.746435(1)  & 1.08773(2) & 0.6228 & 0.1309 & 0.022 & nsO, tdp, {\it sh}, nsx \\ 
OGLE-SMC-CEP-2681  & 2.076572(2)  & 1.29320(2) & 0.6228 & 0.1197 & 0.039 & sh, nsx \\  
                   & 2.076572(2)  & 1.33720(4) & 0.6439 & 0.1197 & 0.024 & \\
OGLE-SMC-CEP-2686  & 1.2856603(7) & 0.785072(7)& 0.6106 & 0.1504 & 0.032 & nsO, cf\\
OGLE-SMC-CEP-2805  & 2.294958(4)  & 1.41812(4) & 0.6179 & 0.1200 & 0.030 & sh, cf\\
OGLE-SMC-CEP-2813  & 2.269260(2)  & 1.40783(2) & 0.6204 & 0.1164 & 0.025 & sh, cf, nsx \\ 
OGLE-SMC-CEP-2860  & 1.4725949(7) & 0.90013(1) & 0.6113 & 0.1455 & 0.019 & al, nsO, nsx \\ 
OGLE-SMC-CEP-2866  & 0.880405(1)  & 0.541328(6)& 0.6149 & 0.1652 & 0.027 & nsO, tdp \\
OGLE-SMC-CEP-2883  & 2.675263(5)  & 1.71600(8) & 0.6414 & 0.0977 & 0.024 & cf \\
OGLE-SMC-CEP-2910  & 2.505797(4)  & 1.60440(8) & 0.6403 & 0.1208 & 0.021 & nsx \\ 
OGLE-SMC-CEP-2941  & 2.951978(3)  & 1.87680(7) & 0.6358 & 0.1216 & 0.016 & nsO, nsx, ap \\ 
OGLE-SMC-CEP-2958  & 1.293953(1)  & 0.78923(2) & 0.6099 & 0.1404 & 0.018 & \\
OGLE-SMC-CEP-2987  & 2.432181(5)  & 1.51720(6) & 0.6238 & 0.1146 & 0.027 & nsO, nsx \\ 
OGLE-SMC-CEP-3033  & 2.710326(5)  & 1.72610(8) & 0.6369 & 0.1090 & 0.024 & \\
OGLE-SMC-CEP-3040  & 2.493781(3)  & 1.59439(4) & 0.6393 & 0.1198 & 0.021 & nsO, {\it sh}, cf\\
                   & 2.493781(3)  & 1.49604(5) & 0.5999 & 0.1198 & 0.017 & \\
OGLE-SMC-CEP-3094  & 2.565406(3)  & 1.63308(4) & 0.6366 & 0.0999 & 0.025 & nsx \\ 
OGLE-SMC-CEP-3098  & 1.366169(1)  & 0.83258(2) & 0.6094 & 0.1582 & 0.021 & al, nsx \\ 
                   & 1.366169(1)  & 0.85752(3) & 0.6277 & 0.1582 & 0.018 & \\
OGLE-SMC-CEP-3143  & 1.0620663(6) & 0.65056(1) & 0.6125 & 0.1872 & 0.017 & \\
OGLE-SMC-CEP-3172  & 1.662515(1)  & 1.03777(2) & 0.6242 & 0.1271 & 0.025 & sh, nsx, nsO \\ 
OGLE-SMC-CEP-3210  & 1.1969483(5) & 0.729788(9)& 0.6097 & 0.1546 & 0.025 & nsO \\
OGLE-SMC-CEP-3239  & 1.796954(2)  & 1.11760(3) & 0.6219 & 0.1446 & 0.031 & nsx, {\it sh} \\ 
OGLE-SMC-CEP-3249  & 1.262314(2)  & 0.76782(2) & 0.6083 & 0.1154 & 0.027 & nsO, tdp\\
OGLE-SMC-CEP-3292  & 2.501456(2)  & 1.54822(4) & 0.6189 & 0.1283 & 0.020 & nsO, cf, ap, {\it sh} \\
OGLE-SMC-CEP-3298  & 2.680077(2)  & 1.71242(5) & 0.6389 & 0.1194 & 0.019 & nsO, sh, cf, nsx \\ 
OGLE-SMC-CEP-3303  & 1.841813(3)  & 1.14780(4) & 0.6232 & 0.1115 & 0.026 & nsO, nsx, tdp, ap \\
OGLE-SMC-CEP-3310  & 1.5201938(9) & 0.92163(2) & 0.6063 & 0.1035 & 0.017 & ? ($S/N=3.79$), nsO \\  
OGLE-SMC-CEP-3312  & 2.117703(2)  & 1.32009(3) & 0.6234 & 0.1227 & 0.024 & sh, nsx \\ 
OGLE-SMC-CEP-3317  & 2.000277(6)  & 1.24893(5) & 0.6244 & 0.1217 & 0.024 & sh, nsO, nsx \\ 
OGLE-SMC-CEP-3319  & 3.088226(7)  & 1.9626(1)  & 0.6355 & 0.0972 & 0.025 & al, nsx \\ 
OGLE-SMC-CEP-3343  & 1.831819(2)  & 1.14484(2) & 0.6250 & 0.1208 & 0.038 & sh, cf, nsx \\ 
OGLE-SMC-CEP-3349  & 1.922025(3)  & 1.20343(3) & 0.6261 & 0.1031 & 0.029 & sh, al, nsx, cf \\ 
OGLE-SMC-CEP-3479  & 1.1644782(8) & 0.71181(1) & 0.6113 & 0.1550 & 0.028 & ap \\ 
OGLE-SMC-CEP-3493  & 1.704786(1)  & 1.03461(2) & 0.6069 & 0.1492 & 0.020 & cf \\
OGLE-SMC-CEP-3541  & 1.3299750(6) & 0.80977(1) & 0.6089 & 0.1558 & 0.018 & nsO, nsx \\ 
OGLE-SMC-CEP-3576  & 1.5859739(8) & 0.96288(1) & 0.6071 & 0.1473 & 0.018 & nsx, cf \\
OGLE-SMC-CEP-3590  & 2.007673(1)  & 1.25295(2) & 0.6241 & 0.1389 & 0.020 & {\it sh}, cf, nsx \\ 
OGLE-SMC-CEP-3624  & 1.4610331(7) & 0.89074(2) & 0.6097 & 0.1541 & 0.014 & ? ($S/N=3.58$)\\
OGLE-SMC-CEP-3668  & 2.738933(3)  & 1.74719(4) & 0.6379 & 0.1203 & 0.020 & nsO, cf, nsx \\ 
OGLE-SMC-CEP-3789  & 1.4240538(7) & 0.86923(1) & 0.6104 & 0.1441 & 0.020 & \\
OGLE-SMC-CEP-3903  & 0.9546276(6) & 0.58594(1) & 0.6138 & 0.1822 & 0.018 & ? ($S/N=3.77$) \\
OGLE-SMC-CEP-3913  & 3.053017(4)  & 1.93793(7) & 0.6348 & 0.1166 & 0.018 & al \\ 
OGLE-SMC-CEP-3944  & 2.162091(2)  & 1.34739(3) & 0.6232 & 0.0989 & 0.028 & sh, nsx, nsO \\
OGLE-SMC-CEP-3977  & 1.542764(3)  & 0.96793(3) & 0.6274 & 0.1187 & 0.044 & al \\
OGLE-SMC-CEP-3987  & 3.368937(9)  & 2.1393(1)  & 0.6350 & 0.1025 & 0.038 & al, {\it sh}, nsO, nsx \\ 
\end{tabular}
\end{table*}

\begin{table*}
\contcaption{}
\begin{tabular}{lrrrrrr}
star & $\Po$ (d) & $\Px$ (d) & $\pxpo$ & $A_{\rm 1O}$ & $A_{\rm x}/A_{\rm 1O}$ & remarks \\
\hline
OGLE-SMC-CEP-4011  & 1.814599(2)  & 1.13516(3) & 0.6256 & 0.1368 & 0.025 & sh, nsx, cf \\ 
OGLE-SMC-CEP-4046  & 2.248845(4)  & 1.39808(4) & 0.6217 & 0.1210 & 0.036 & sh, nsx, cf \\
OGLE-SMC-CEP-4061  & 2.573926(3)  & 1.64745(6) & 0.6401 & 0.1125 & 0.018 & nsx, nsO \\
OGLE-SMC-CEP-4068  & 2.010614(2)  & 1.24762(2) & 0.6205 & 0.1231 & 0.028 & sh, nsO, nsx \\
OGLE-SMC-CEP-4157  & 1.256266(1)  & 0.76187(3) & 0.6065 & 0.1485 & 0.021 & ? ($S/N=3.34$)\\
OGLE-SMC-CEP-4188  & 1.520138(1)  & 0.92475(2) & 0.6083 & 0.1482 & 0.021 & ap \\
OGLE-SMC-CEP-4205  & 1.944865(3)  & 1.21286(2) & 0.6236 & 0.1208 & 0.049 & sh, nsx, cf \\
OGLE-SMC-CEP-4232  & 1.721685(2)  & 1.07401(5) & 0.6238 & 0.1536 & 0.015 & ? ($S/N=3.76$)\\
OGLE-SMC-CEP-4250  & 1.466489(1)  & 0.89468(3) & 0.6101 & 0.1502 & 0.021 & \\
OGLE-SMC-CEP-4255  & 2.091244(5)  & 1.30621(5) & 0.6246 & 0.0988 & 0.038 & sh, al, nsx \\
OGLE-SMC-CEP-4262  & 2.093166(6)  & 1.30318(5) & 0.6226 & 0.0885 & 0.049 & sh, nsx \\ 
OGLE-SMC-CEP-4303  & 1.1679163(9) & 0.71344(1) & 0.6109 & 0.1487 & 0.030 & al, nsO, nsx \\ 
OGLE-SMC-CEP-4316  & 1.295785(1)  & 0.78725(1) & 0.6076 & 0.1686 & 0.029 & \\
OGLE-SMC-CEP-4378  & 2.67350(2)   & 1.7078(1)  & 0.6388 & 0.1035 & 0.030 & nsO, tdp, al\\
OGLE-SMC-CEP-4388  & 2.034294(5)  & 1.26659(5) & 0.6226 & 0.1024 & 0.044 & sh \\
OGLE-SMC-CEP-4394  & 1.095018(1)  & 0.67156(1) & 0.6133 & 0.1309 & 0.032 & nsO \\
OGLE-SMC-CEP-4395  & 2.869590(7)  & 1.8288(1)  & 0.6373 & 0.1110 & 0.030 & {\it sh}, nsx \\
OGLE-SMC-CEP-4462  & 2.136558(4)  & 1.32484(6) & 0.6201 & 0.1031 & 0.029 & sh, nsx \\ 
OGLE-SMC-CEP-4587  & 1.1416708(7) & 0.69856(1) & 0.6119 & 0.1761 & 0.024 & \\
OGLE-SMC-CEP-4627  & 3.17304(2)   & 1.95769(5) & 0.6170 & 0.0735 & 0.105 & cf, ap \\ 
\hline
\end{tabular}
\end{table*}

\section{Comparison with Soszy\'nski et al. (2010)}\label{ssec:igor}
For each star \cite{ogle_cep_smc} (S10 in the following) provided two periods, $\Po$ and $\Px$, and these are the only quantities we can compare. For the majority of stars there is an excellent agreement between our studies. In 6 stars the small differences are a consequence of the complex form in which additional variability appears in the frequency spectrum (see Fig.~\ref{fig:ilu} and Sect.~\ref{ssec:overview}) -- two close peaks of similar height or cluster of peaks. Small differences in the analysis (detrending, outlier rejection) may slightly alter the relative height of the peaks and lead to slightly different frequencies adopted in the two studies. The difference in the resulting period ratios is negligible however. Two examples are illustrated in the top two panels of Fig.~\ref{fig:ilu} in which filled diamonds mark the frequency adopted in this study and open diamonds mark the frequency adopted in S10. 

More significant differences are due to alias ambiguities and are found for six stars, described below on a star-by-star basis. Still, only for one star the period ratios in the two studies differ by 3\thinspace per cent. For five other stars the difference is less than 1\thinspace per cent.

OGLE-SMC-CEP-1127 -- the difference for this star is a result of long-term trend, likely not removed by S10. Its one day alias falls at $P/\Po\approx 0.6084$ and was interpreted as additional variability in S10. The trend is well visible in the residual data. After removing the trend with third order polynomial the one day alias also disappears, but another signal at $\pxpo\approx 0.6270$ is present, which we interpret as due to additional variability.

OGLE-SMC-CEP-2131 -- the difference for this star is a result of strongly non-stationarity of the first overtone. One-day alias (at $P/\Po\approx 0.6215$) of the strong, residual peak at the frequency of first overtone was likely interpreted as additional variability by S10. After time-dependent prewhitening the ambiguous signal disappears, but we detect another significant peak nearby, at $\pxpo\approx 0.6163$, and interpret as due to additional variability.

OGLE-SMC-CEP-3094 -- after prewhitening with $\fo$ and its harmonics (and detrending) the strongest peak is present at $\pxpo=0.6366$, as given in Tab.~\ref{tab:tab}. Its daily alias falls very close to $\fo$ (resolved) and likely was prewhitened by S10 first. Then, the highest peak is found at $\pxpo\approx0.6327$ as is reported in S10. Both solutions are likely.

OGLE-SMC-CEP-3098 -- after prewhitening with $\fo$ and its harmonics (and detrending) the strongest peak in the $P/\Po\in(0.6,\,0.65)$ range is detected at $\pxpo\approx 0.6094$ as reported in Tab.~\ref{tab:tab}. Its daily alias is higher however, it falls at $P/P_1\approx 0.913$ and can be interpreted as other periodicity. In S10 this peak was likely prewhitened first. Then, the highest peak in the frequency range of interest appears at $\pxpo\approx0.6101$ as reported in S10. Both solutions are likely.

OGLE-SMC-CEP-3349 -- after prewhitening with $\fo$ and its harmonics (and detrending) the highest peak in the frequency range of interest appears at $\pxpo\approx 0.6261$, as reported in Tab.~\ref{tab:tab}. Its daily alias is higher however. There are two lower peaks around; one of them at $\pxpo\approx0.6217$ was reported in S10 (interestingly its daily alias is also higher).

OGLE-SMC-CEP-4255 -- after prewhitening with $\fo$ and its harmonics (and detrending) the strongest peak in the $P/\Po\in(0.6,\,0.65)$ range is detected at $\pxpo\approx 0.6246$ ($S/N=5.16$) as given in Tab~\ref{tab:tab}. Its daily alias is higher however, it falls at $P_1/P=0.496$ ($S/N=5.29$). In S10 this peak was likely prewhitened first. Then, the highest peak in the frequency range of interest appears at $\pxpo\approx 0.6252$ as reported in S10. Both solutions are likely.

\begin{figure}
\centering
\resizebox{.95\hsize}{!}{\includegraphics{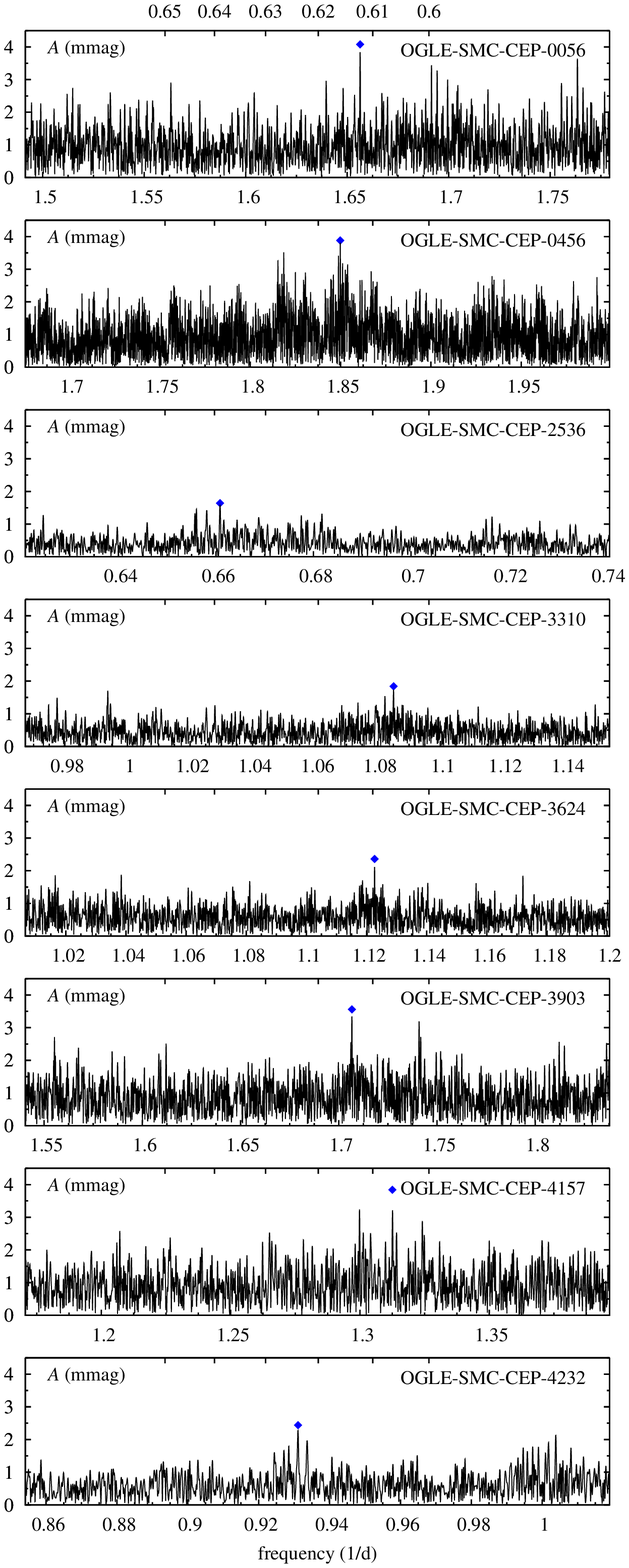}}
\caption{Frequency spectra for eight stars in which detection of the additional variability is weak, with $3.0<S/N<4.0$. Diamond marks the location of the peak and is placed exactly at $S/N=4.0$. Ticks at the top axis of each panel correspond to $P/\Po$ scale, explicitly given at the top of the Figure.}
\label{fig:nothing}
\end{figure}

In eight stars the highest peak in the frequency range of interest corresponds to $\pxpo$ falling well within one of the three sequences visible in Fig.~\ref{fig:pet} (stars marked with triangles), but the $S/N$ is below $4.0$ (it is reported in the last column of Tab~\ref{tab:tab}; these stars also have `?' in remarks column). Frequency spectra for these stars are plotted in Fig.~\ref{fig:nothing}. Likely the additional variability is present in these stars but its amplitude is small as compared to the noise level.

\clearpage
\bsp	
\label{lastpage}
\end{document}